\documentclass[12pt,epsf]{article}

\newcommand{\be}{\begin{equation}}
\newcommand{\ee}{\end{equation}}
\newcommand{\bea}{\begin{eqnarray}}
\newcommand{\eea}{\end{eqnarray}}
\newcommand{\beas}{\begin{eqnarray*}}
\newcommand{\eeas}{\end{eqnarray*}}
\newcommand{\ba}{\begin{array}}
\newcommand{\ea}{\end{array}}

\newcommand{\nn}{\nonumber}

\renewcommand{\b}{\beta}

\newcommand{\w}{\omega}

\renewcommand{\d}{\partial}

\renewcommand{\(}{\left(}
\renewcommand{\)}{\right)}
\newcommand{\tr}{\mathrm{Tr}}
\newcommand{\STr}{\mathrm{STr}}
\newcommand{\sym}{\mathrm{Sym}}
\newcommand{\diag}{\mathrm{diag}}

\newcommand{\text}[1]{{\rm #1}}

\newcommand{\nbox}{{\,\lower0.9pt\vbox{\hrule \hbox{\vrule height 0.2 cm \hskip 0.19 cm \vrule height 0.2 cm}\hrule}\,}}
\newcommand{\Tr}{\ {\rm Tr}\ }
\renewcommand{\v}[1]{\vec{#1}}

\newcommand{\dddot}[1]{\stackrel{\dots}{#1}}

\def\href#1#2{#2}

\begin{document}
\begin{titlepage}
\hfill
\vbox{
    \halign{#\hfil         \cr
           hep-th/0509026 \cr
           } % end of \halign
      }  % end of \vbox
\vspace*{20mm}
\begin{center}
{\Large \bf Poincar\'{e} Invariance in Multiple D-brane Actions}

\vspace*{15mm}
\vspace*{1mm}
%{Authors}
{Dominic Brecher, Paul Koerber, Henry Ling, Mark Van Raamsdonk}

\vspace*{1cm}

{Department of Physics and Astronomy,
University of British Columbia\\
6224 Agricultural Road,
Vancouver, B.C., V6T 1W9, Canada}

\vspace*{1cm}
%%\maketitle
\end{center}

\begin{abstract}
We show that the requirement of Poincar\'{e} invariance (more specifically invariance under boosts/rotations that mix brane directions with transverse directions) places severe constraints on the form of actions describing multiple D-branes, determining an infinite series of correction terms to the currently known actions. For the case of D0-branes, we argue that up to field redefinitions, there is a unique Lorentz transformation rule for the coordinate matrices consistent with the Poincar\'{e} algebra. We characterize all independent Poincar\'{e} invariant structures by describing the leading term of each and providing an implicit construction of a Lorentz invariant completion.
Our construction employs new matrix-valued Lorentz covariant objects built from the coordinate matrices, which transform simply under the (extremely complicated) Lorentz transformation rule for the matrix coordinates.

\end{abstract}

\end{titlepage}

\vskip 1cm
\section{Introduction}

Consider a collection of $N$ D0-branes in flat $R^{d+1}$.\footnote{These could be the usual D0-branes of type IIA string theory with $d=9$, or any other pointlike D-branes arising from higher dimensional branes wrapped on cycles in a compactification.} Their low-energy configurations are described by $d$ $N \times N$ Hermitian matrices $X^i(t)$ \cite{witten}, and their dynamics is controlled by some effective action\footnote{Here, all bulk fields have been set to zero.}
\[
S[X(t)] = \int dt \; L(X(t)) \; .
\]
Now consider the same system as described by an observer in an infinitesimally boosted frame. Again there will be a description in terms of $d$ $N \times N$ Hermitian matrices, related to those in the original frame by a Lorentz transformation
\be
\label{lt}
\tilde{X}^i(t) = \Phi^i[X(t), \beta^j] \; ,
\ee
where $\beta^j$ is the velocity of the boosted frame.
The second observer should describe physics by the same action since the background is unchanged. Since $\tilde{X}$ and $X$ describe physically equivalent configurations, it must be that
\be
\label{invar0}
S[X(t)] = S[\tilde{X}(t)]
\ee
for any $X$.

The goal of this paper will be to understand the Lorentz transformation rule (\ref{lt}) for the matrix coordinates of D-branes, and to understand the constraints imposed on the action by requiring invariance (\ref{invar0}) under this transformation.

The problem we consider here is quite nontrivial\footnote{Indeed, to our knowledge {\it none} of the actions for multiple D-branes that have appeared previously in the literature are Poincar\'{e} invariant apart from the cases $p = -1$ and $p=9$ which are trivial.} because the description of a single brane in terms of covariant embedding coordinates $x^\mu(\tau)$, upon which Lorentz transformations act simply, cannot as yet be generalized to describe multiple branes. In the case of multiple branes, the usual description generalizes the static gauge description of a single brane, where reparametrization invariance is used to set $x^0(\tau)=\tau$ and the space-time embedding is then completely specified by the spatial coordinates $x^i(\tau)$. In this picture, Lorentz transformations which mix space and time are somewhat messy even in the abelian case, and turn out to be extremely complicated for the non-abelian case.

\subsubsection*{Background}

Our analysis here is another step in a program to understand the implementation of and constraints arising from the full set of
bulk space-time symmetries on multiple D-brane actions.\footnote{Early work on understanding the structure of non-abelian D-brane actions based on general principles was initiated by Douglas \cite{douglas1, douglas2}. For general reviews discussing the physics of multiple D-branes, including many additional references, see \cite{myers, taylor}.} The motivation is both to understand the actions themselves, but more generally to understand whether there is some natural geometric language (as we have in the abelian case) in terms of which multiple D-brane actions are simple.

Previously (\cite{old}, see also \cite{dbs}), we considered the case of spatial
diffeomorphisms in a system of D0-branes coupled to gravity. We saw that the matrices describing D0-brane configurations have a very
complicated transformation rule under the diffeomorphisms, and that demanding invariance of the action under such transformations
imposes severe constraints. Most interestingly, we showed the existence\footnote{Here, the existence of a consistent transformation
rule for the coordinate matrices was assumed.} of a covariant matrix-valued vector field built from the coordinate matrices,
which allowed us to construct (though somewhat implicitly) the most general action invariant under spatial diffeomorphisms.

The restriction to spatial diffeomorphisms in our previous work was made specifically to avoid transformations that mix world-volume directions with transverse directions described by matrices. As we have seen above, such transformations present an additional complication, since they look complicated even in the abelian case if we restrict to the static gauge. Before attempting to analyze the full group of space-time diffeomorphisms, it is natural to begin with the simplest case for which the additional complication arises, namely Lorentz transformations for a system of D0-branes in flat space. This is the focus of the present paper.

\subsubsection*{Outline and summary}

We begin in section 2 with an order-by-order analysis of the transformation law.
We determine the infinitesimal Poincar\'{e} transformations for a single particle in static gauge,
and show that the simplest generalization of these to the matrix case does not respect the Poincar\'{e} algebra.
We find that it is possible, working to all orders in $X$ and up to two commutators, to add commutator terms to the boost transformation rule such that the Poincar\'{e} algebra is restored. The success of this procedure is highly nontrivial and provides significant evidence that a consistent transformation rule exists to all orders.\footnote{Of course, this should be guaranteed if string theory is consistent and Lorentz invariant in flat space.} Assuming this, we show that the boost transformation law is unique up to field redefinitions which do not affect the other Poincar\'{e} transformations.

In section 3, we begin our analysis of the invariant actions, now working order-by-order in $X$ in the static gauge. Using the transformation rule from section 2, we find that it is possible to add terms order-by-order to the leading $\tr(\dot{X}^2)$ kinetic term and to the simplest potential term $\tr[X^i,X^j]^2$ to obtain (independent) Lorentz invariant results (we work up to order $X^6$). We determine a necessary condition that must be satisfied by the leading term of any Poincar\'{e} invariant structure, generalizing the necessary conditions of  time-reversal and Galilean invariance (including parity) in the abelian case. Finally, we show that for any choice of field there is at most one invariant action depending on $\dot{X}$ and not $X$ or higher derivatives of $X$. Unfortunately, we find that any invariant generalization of the abelian kinetic term may be written in this way for some appropriate choice of field, so it is not clear whether a canonical non-abelian generalization of the usual relativistic kinetic term exists.

In section 4, we look for a more natural way to write Lorentz invariant actions. As in our previous studies, we look for matrix-valued covariant objects defined as fields over space-time from which we can build manifestly invariant actions as integrals over space-time. We find (at least up to fifth order in $X$) that there exists a covariant matrix vector field $V^\mu(y)$ built from $X$ but transforming simply under a Lorentz transformation $\tilde{y}^\mu=\Lambda^\mu{}_\nu\, y^\nu$ as
\[
\tilde{V}^\mu(\tilde{y}) = \Lambda^\mu{}_\nu V^\nu(y) \; .
\]
In the abelian case, $V$ is the derivative of the proper distance to the trajectory along a geodesic which intersects the trajectory orthogonally. In addition, we find a covariant matrix distribution function $\Theta(y)$ which reduces in the abelian case to
\[
\Theta(y) = \int d\tau\; \sqrt{-\d_\tau x^\mu \d_\tau x_\mu }  \ \delta^{d+1}(x^\nu(\tau)-y^\nu) \; .
\]

In section 5, we show that all Poincar\'{e} invariant actions may be written using these two covariant objects as
\[
S = \int \; d^{d+1} y \; \tr({\cal L}(V(y))\Theta(y)) \; ,
\]
where ${\cal L}$ is a scalar built from $V$. The independent Poincar\'{e} invariant structures may be characterized by their leading terms, which may either be $\int dt\  \tr(\dot{X}^2)$ or may be written as the integral of a Lagrangian $L(X, \dot{X}, \ddot{X}, \dots)$ with an even number of $X$s and time derivatives satisfying
\[
\partial_\epsilon L(X + \epsilon, \dot{X}, \dots) = \partial_\beta L(X, \dot{X} + \beta, \dots)=0 \; ,
\]
i.e.\ a term with all $X$s and $\dot{X}s$ appearing in complete commutators. This is precisely the necessary condition we found in section 3, so we conclude that the one-to-one correspondence between Poincar\'{e} invariant structures and Galilean (and time-reversal) invariant leading terms familiar from the abelian case extends to the non-abelian case also.

In section 6, we discuss the couplings to space-time supergravity
fields. We note that Lorentz symmetry also implies higher order
corrections to these terms, and in particular to the various
conserved space-time currents associated with the branes (e.g. the
stress-energy tensor or D$p$-brane currents).

We offer a few concluding remarks in section 7.

\subsubsection*{Relation to other work}

Most previous work on corrections to flat-space non-abelian D-brane actions focuses on the gauge field on the world-volume of the D$p$-branes. It is now known (see references below) that these corrections do not in general take the simple symmetrized form of \cite{tseytlin}, but contain (covariant) derivatives and commutators.  Using T-duality, one should be able to obtain D0-brane actions from these D$p$-brane actions that are consistent with the requirement of Poincar\'{e} invariance.
However, we have not attempted to verify if this is true; any attempts to do so should take into account field redefinitions that change the form of the Lorentz transformation law.

Corrections to the non-abelian D$p$-brane gauge field effective action were calculated via direct string amplitude calculation
\cite{Barreiro:2005hv,Oprisa:2005wu}, requiring supersymmetry \cite{Collinucci:2002ac,Drummond:2003ex,Grasso:2004mk},
requiring the existence of BPS solutions \cite{Koerber:2002zb}, the Seiberg-Witten map \cite{Bordalo:2004xg} and the spectrum of intersecting branes
\cite{Hashimoto:1997gm,Sevrin:2003vs}. Derivative corrections to the abelian effective action were found in \cite{Bachas:1999um} (for
the transverse scalars) and in \cite{Wyllard:2000qe,deRoo:2003xv} (for the gauge field). For a more complete set of references, see \cite{Koerber:2004ze}.

\section{Poincar\'{e} transformation rules for multiple D0-branes}

The usual description of the low-energy degrees of freedom for a
collection of $N$ D0-branes utilizes one $N \times N$ Hermitian matrix
$X^i(t)$ for each spatial direction, each a function of the world-volume time. In this section, we would like to understand how these matrix coordinates transform under Poincar\'{e} transformations.

\subsubsection*{Transformation rules in the abelian case}

We begin by recalling the Poincar\'{e} transformation rules for a single
D0-brane in $(d+1)$-dimensional Minkowski space. In this case, the
brane can be described by a set of embedding functions
$x^\mu(\tau)$ whose Poincar\'{e} transformations are simply
\be
\label{brtran}
\tilde{x}^\mu(\tau)=\Lambda^\mu{}_\nu\ x^\nu(\tau) + a^\mu \; .
\ee
To ensure that the system has the correct number of physical degrees of freedom, we must also demand the (gauge) invariance of the action
under world-volume diffeomorphisms
\be
\label{repar}
\tau \rightarrow \tau'=f(\tau) \; .
\ee

It would be nice if the transformation rules (\ref{brtran}) could
be extended in some simple way to the non-abelian case, e.g. by
introducing matrices for all space-time (rather than just
spatial) directions. Unfortunately, there is no obvious way to do
this, and such a description would seem to require
an analogue of the world-volume reparametrization symmetry capable of
eliminating an entire matrix worth of degrees of freedom.

\subsubsection*{Abelian transformation rules: static gauge}

The description that does generalize easily to the non-abelian case
is one in which the world-volume reparametrization invariance has
been fixed by choosing \be \label{static} x^0(\tau) = \tau \; . \ee
To make progress in understanding the non-abelian transformation
law, we will therefore rewrite the abelian transformation rules in
this `static gauge' description and try to generalize these to the
case of multiple branes.

Throughout, we will use $t$ to denote the world-volume parameter in static gauge, to emphasize that world-volume time has been set equal to target-space time.  Now starting from a set of static gauge embedding functions $x^\mu(t)=(t,x^i(t))$, it is
evident that applying the coordinate transformation (\ref{brtran})
will bring us out of the static gauge. Therefore, we must
combine the transformation (\ref{brtran}) with a compensating
world-volume diffeomorphism (\ref{repar}) which restores the static
gauge.
The resulting static-gauge Poincar\'{e} transformation rule is
\be
\tilde{x}^i(t) = \Lambda^i{}_0\  h^{-1}(t) + \Lambda^i{}_j\
x^j(h^{-1}(t)) + a^i,
\ee
where $h(t) \equiv \Lambda^0{}_0\ t + \Lambda^0{}_i\ x^i(t) + a^0$.
While this is much more complicated than (\ref{brtran}), it acts
only on the spatial coordinates, and therefore has a chance of
generalizing to the non-abelian case.

In order to simplify matters as much as possible, we specialize to
the case of infinitesimal transformations.  Then the static gauge
transformation rules for translations, time translations, rotations,
and boosts take the form \bea \delta_{\vec{a}} x^i &=& a^i \; , \cr
\delta_{a^0} x^i &=& -a^0 \dot{x}^i \; , \cr \delta_\w x^i &=&
\w^{ij} x^j \; ,\cr \delta_\b x^i &=& \b^i t - \b^j \dot{x}^i x^j \;
. \label{abelrules} \eea It is the non-linearity in the
infinitesimal transformation rule for boosts that makes a
generalization to the non-abelian case quite nontrivial.

\subsubsection*{The Poincar\'{e} algebra as a consistency condition}

The Poincar\'{e} transformation rules for the matrix coordinates of
multiple D0-branes should be some generalization of
(\ref{abelrules}). Since they must reduce to (\ref{abelrules}) in the
case where all matrices are diagonal, it must be that
all corrections involve commutators of matrices. A further constraint
comes from demanding that the Poincar\'{e} algebra is still satisfied by
the non-abelian transformations.

The rotation, translation, and time translation rules in
(\ref{abelrules}) are linear in $X$ and generalize unambiguously to
the non-abelian case without modification. We will assume that these receive no commutator corrections, since it is consistent with the algebra of rotations and translations (and certainly very natural) to do so.

For the boost transformation law in (\ref{abelrules}), an ordering issue arises since there are various non-abelian generalizations of the quadratic term. The Poincar\'{e} algebra demands that the correct generalization must satisfy
%\begin{subequations}
\bea
\label{algebrabb}
(\delta_{\tilde{\beta}} \delta_\beta - \delta_{\tilde{\beta}}
\delta_\beta)X &=& \delta_{\omega^{ij} = \beta^i \tilde{\beta}^j -
  \beta^j \tilde{\beta}^i} X  \; , \\
\label{algebraab}
(\delta_{\vec{a}} \delta_\beta - \delta_\beta
\delta_{\vec{a}}) X &=& \delta_{a^0 = \beta \cdot a} X \; , \\
\label{algebra0b}
(\delta_{a^0} \delta_\beta - \delta_\beta
\delta_{a^0}) X &=& \delta_{\vec{a} = a^0 \vec{\beta}} X \; , \\
\label{algebrabw}
(\delta_\beta \delta_\omega - \delta_\omega
\delta_\beta)X &=& \delta_{\beta^i = \omega^{ij} \beta^j} X \; .
\eea
%\end{subequations}
In fact, it is easy to show that (\ref{algebrabb}) is not satisfied for {\it any} of the possible orderings of the quadratic term without adding corrections to the transformation law at higher orders in $X$.

We will therefore write the putative Poincar\'{e} transformation rules for the non-abelian case as
\bea
\delta_{\vec{a}} X^i &=& a^i \; , \cr
\delta_{a^0} X^i &=& -a^0 \dot{X}^i \; , \cr
\delta_\w X^i &=& \w^{ij}X^j \; , \cr
\delta_\b X^i &=& \b^i t - \b^j \sym(\dot{X}^i X^j) + \beta^j T^{ij} \; .
\label{nonabelrules}
\eea
where $\sym$ indicates the symmetrized ordering ($\sym(AB) = {1 \over 2} (AB + BA)$) and $T^{ij}$ stands for some series of terms that vanish for diagonal $X^i$.\footnote{Note that we are not assuming that the quadratic term is symmetric, since $T^{ij}$ may contain quadratic commutator terms.} We now ask whether it is possible to choose Hermitian $T^{ij}$ built from $X$ and its derivatives such that the constraints (\ref{algebrabb},\ref{algebraab},\ref{algebra0b},\ref{algebrabw}) are satisfied.

First, the constraint (\ref{algebrabw}) is satisfied automatically as long as $T^{ij}$ is a tensor under the rotation group. Any tensor built from the vector $X^i$ and its derivatives satisfies this constraint.

Next, the constraint (\ref{algebra0b}) is satisfied as long as $T^{ij}$ contains no explicit time dependence.

The constraint (\ref{algebraab}) implies that
\[
\partial_\epsilon T^{ij} (X + \epsilon) = 0 \; ,
\]
i.e.\ all undifferentiated $X$s must appear in commutators.

Finally, the constraint (\ref{algebrabb}) implies that
\be
\label{finconstr}
\beta^j \delta_{\tilde{\beta}} S^{ij} - \tilde{\beta}^j \delta_{\beta} S^{ij} = \beta^i \tilde{\beta}^j X^j - \tilde{\beta}^i \beta^j X^j \; ,
\ee
where
\[
S^{ij} = -\sym(\dot{X}^i X^j) + T^{ij} \; .
\]
This turns out to be quite nontrivial, and we resort to an order-by order approach to check whether a solution exists.

\subsubsection*{Order-by-order solution}
\label{orderbyorder}

It is straightforward to check that (\ref{finconstr}) holds at leading orders with $T^{ij} = 0$, but breaks down at order $X^3$ unless we add commutator corrections $T^{ij}$ at order $X^4$. We find that these must satisfy
\be
\label{fourth}
\delta_\beta^0 T^{ij}_{(4)} = \frac{1}{8}\b^k \left(  - [\ddot{X}^i,[X^k,X^j]] + [\dot{X}^k,[X^j,\dot{X}^i]] - [\dot{X}^j,[X^k,\dot{X}^i]]  \right) \; ,
\ee
where $\delta_\beta^0$ indicates the variation keeping only the order $X^0$ term in the boost transformation law (\ref{nonabelrules}).

Appropriate corrections are possible at this order, for example\footnote{The operation $\sym$ is assumed to treat commutator expressions as a unit in the symmetrization.}
\be
\label{firstnon}
T^{ij}_{(4)} = \frac{1}{8} \sym \left(  -\dot{X}^k [\ddot{X}^i,[X^k,X^j]] + \dot{X}^k [\dot{X}^k,[X^j,\dot{X}^i]] - \dot{X}^k [\dot{X}^j,[X^k,\dot{X}^i]]  \right) \; .
\ee
Note that the equations (\ref{fourth}) determining $T^{ij}$ at this order are overconstrained in the sense that solutions exist only for special choices of the right-hand side. Thus, the existence of a solution can be taken as a first piece of evidence that the Poincar\'{e} transformation rules admit an extension to the non-abelian case. This expression is not unique, but we will see shortly that all possible solutions are related by a class of field redefinitions.

We might now proceed ad nauseam checking at each order in $X$ that a
choice of $T^{ij}$ exists such that the constraint (\ref{finconstr})
is satisfied to the appropriate order. Instead, we will take a
slightly more refined approach that we now describe.

\subsubsection*{The expansion in number of commutators}

Throughout this paper it turns out to be possible to obtain partial all-order results in powers of $X$, when expanding in the number of commutators. Suppose we have a term consisting of a product of matrices at some order. Then we can always symmetrize this product and compensate by adding appropriate terms with commutators. These extra terms can in turn be symmetrized, where the commutators are considered as a unit under the symmetrization, by adding terms with more commutators and so on. In the end, one obtains a sum of symmetrized products. Because of the overall symmetrization the number of commutators in a term has a definite meaning. Only the first term does not contain
commutators and remains in the abelian limit. The other terms are non-abelian corrections with a fixed number of
commutators. If the non-abelian corrections are small it would be sensible to calculate only up to a certain number of commutators, and we will often employ such an expansion in this paper.

Using this approach, we have checked that a solution to the constraint (\ref{finconstr}) and therefore a consistent boost transformation rule exists to all orders in $X$ at second order in commutators. The result, derived in appendix \ref{2commtranslawcalc} is
\bea
\lefteqn{T^{il}=} \nn \\
& & \sym \Bigg[\frac{1}{8} E^{jt} \dot{X}^t (- [\ddot{X}^i,[X^j,X^l]] + [\dot{X}^j,[X^l,\dot{X}^i]] - [\dot{X}^l,[X^j,\dot{X}^i]]) \cr
&& -\frac{1}{8} E^{j_1t_1} \dot{X}^{t_1} E^{j_2t_2} \dot{X}^{t_2}
\Big(\frac{1}{3} \dddot{X}{}^i [X^{j_1},[X^{j_2},X^l]] \nn \\
& & +\ddot{X}^{j_1}[X^{j_2},[\dot{X}^i,X^l]]
+\ddot{X}^{i}[\dot{X}^{j_1},[X^{j_2},X^l]]\Big) \nn \\
& & -\frac{1}{8} E^{j_1t_1} \dot{X}^{t_1} E^{j_2t_2} \dot{X}^{t_2} E^{j_3t_3} \dot{X}^{t_3}
\ddot{X}^i \ddot{X}^{j_1} [X^{j_2},[X^{j_3},X^l]] \nn\\
& & -\frac{1}{8} E^{j_1t_1} \dot{X}^{t_1} E^{j_2t_2} \dot{X}^{t_2}\([\ddot{X}^i,X^{j_1}][X^{j_2},\dot{X}^l]+[\dot{X}^i,X^{j_1}][\dot{X}^{j_2},\dot{X}^l]
+[\dot{X}_i,\dot{X}_{l}][\dot{X}_{j_1},{X}_{j_2}]\) \nn\\
& & +\frac{1}{8} E^{j_1t_1} \dot{X}^{t_1} E^{j_2t_2} \dot{X}^{t_2} E^{j_3t_3} \dot{X}^{t_3} \Big( \frac{1}{3} \dddot{X}{}^i [X^l,X^{j_1}][X^{j_2},\dot{X}^{j_3}]
- \ddot{X}^i [\dot{X}^l,X^{j_1}][X^{j_2},\dot{X}^{j_3}] \nn \\
& & + \ddot{X}^{j_1} [\ddot{X}^i,X^{j_2}][X^{j_3},X^{l}]
- \ddot{X}^l [\dot{X}^i,X^{j_1}][\dot{X}^{j_2},X^{j_3}]\Big) \nn \\
& & - \frac{1}{8} E^{j_1t_1} \dot{X}^{t_1} E^{j_2t_2} \dot{X}^{t_2} E^{j_3t_3} \dot{X}^{t_3} E^{j_4t_4} \dot{X}^{t_4}\ddot{X}^i \ddot{X}^{j_1}
[X^{j_2},\dot{X}^{j_3}][X^{j_4},X^l] \Bigg]  \, ,
\label{2commtranslawterm}
\eea
where $E^{jt}$ is the inverse tensor
\be
\label{Einverse}
E^{jt}\(\delta^{tp}-\dot{X}^t \dot{X}^p\)=\delta^{jp}.
\ee
The existence of a full solution even to second order in commutators is extremely nontrivial and suggests strongly that a consistent boost transformation law exists to all orders. While we are not able to prove this, we will now show that any such transformation rule must be unique up to a class of field redefinitions.

\subsubsection*{Uniqueness of the transformation rule up to field redefinitions}

It is easy to see that we cannot expect a completely unique solution to the constraints outlined so far for the non-abelian generalization of the boost transformation law. For, consider a new variable
\be
\label{FR}
\tilde{X}^i = X^i + F^i(X) \; ,
\ee
where $F$ is a polynomial in $X$ (possibly infinite) defined so that $\tilde{X}$ and $X$ agree in the abelian case and $\tilde{X}$ has the same transformation rule as $X$ under rotations, translations, and time translations. These will be true as long as
\begin{itemize}
\item
$F^i(X)$ is a vector built from $X$ and its derivatives that vanishes for diagonal $X$;
\item
$F^i(X)$ has no explicit time dependence;
\item
$F^i(X)$ is translation invariant (has all undifferentiated $X$s appearing in commutators).
\end{itemize}
The transformation rule for $\tilde{X}$
under boosts (obtained by transforming the right side of (\ref{FR})
and rewriting all occurrences of $X$ in terms of $\tilde{X}$ by
inverting (\ref{FR})) will generally be different from that of $X$,
with the lowest order change in $T^{ij}$ given by \be
\label{FRlowestorder} \beta^j \Delta T^{ij} = \delta^0_\beta F^i \;
. \ee As above, $\delta^0_{\b} X^i=\b^i t$ denotes the order $X^0$
term in the boost transformation law. The new transformation rule
will necessarily be consistent with the Poincar\'{e} algebra, as
this follows directly from consistency of the transformation rule
for $X$. Since the other Poincar\'{e} transformations remain the
same, the boost transformation rule for $\tilde{X}$ represents a new
solution to the constraint (\ref{finconstr}).

On the other hand, it is straightforward to show that all nonuniqueness in the transformation law may be associated with such field redefinitions. For suppose that there exist two different transformation laws $\delta_\beta X$ and $\tilde{\delta}_\beta X$ for which the constraints (\ref{finconstr}) and all other constraints of that subsection are satisfied. Then the leading order difference $\Delta_0 T^{ij}$ between  $T^{ij}$ and $\tilde{T}^{ij}$ must satisfy
\be
\beta_2^j \delta^0_{\b_1} \Delta_0 T^{ij} - \beta_1^j \delta^0_{\b_2} \Delta_0 T^{ij}=0 \, .
\label{extra}
\ee
It follows that
$\Delta_0 T^{ij}$ is of the form\footnote{Here, the round brackets denote symmetrization.}
\be
\Delta_0 T^{ij} = \sym \(\dot{X}^{k_1} \cdots \dot{X}^{k_m} D^{i(jk_1\ldots k_m)}\),
\ee
where $D^{i(jk_1\ldots k_m)}$ is an arbitrary tensor that cannot contain $X$ and $\dot{X}$ outside of
commutators. But (\ref{FRlowestorder}) shows that this is the same leading order difference that arises in making a field redefinition (\ref{FR}) with
\be
\label{fchoice}
F^i = \frac{1}{m+1} \, \sym \(\dot{X}^j \dot{X}^{k_1} \cdots \dot{X}^{k_m} D^{i(jk_1\ldots k_m)}\) \; .
\ee
If $\delta_\beta X$ and $\tilde{\delta}_\beta X$ differed at order $X^n$, then $\delta_\beta \tilde{X}$ and $\tilde{\delta}_\beta X$ may differ
only at higher order. We may then repeat our procedure, making a further field redefinition to remove the leading discrepancy at this order,
and so forth, so that after an infinite number of steps we find some new variable $X^i_\infty$ such that $\delta_\beta X_\infty$ is the same as $\tilde{\delta}_\beta X$. Note that the $F$ in (\ref{fchoice}) satisfies all the constraints of the previous paragraph, since the discussion before equation (\ref{finconstr}) implies that $\Delta_0 T^{ij}$ should satisfy these same constraints.

Thus, any two consistent generalizations of the Poincar\'{e} transformations to the non-abelian case are related by a field redefinition that is trivial in the abelian case and preserves the transformation rules for rotations, translations, and time translations.

\section{Poincar\'{e} invariant actions for multiple D0-branes}

In this section, we begin to investigate the constraints imposed by Poincar\'{e} invariance on the form of the effective action. We will assume henceforth that a consistent boost transformation rule exists (generalizing (\ref{2commtranslawterm}) to all orders). In particular, we assume that (as in (\ref{2commtranslawterm})) it is possible to write such a transformation law without introducing any dimensionful coefficients, such that all terms will have one less time derivative than the number of $X$s.\footnote{In the context of string theory one might wonder if the correct transformation law involves higher order terms with explicit powers of $\alpha'$. However, our results from the previous section suggest that there does exist a valid transformation law without any $\alpha'$ dependence and that any $\alpha'$ dependent transformation law should be equivalent to this by a field redefinition (that would necessarily involve explicit factors of $\alpha'$).}

In our discussions below, we consider explicitly only single-trace actions, which arise in string theory at the leading order in $g_s$, but we expect that most of the results generalize readily to the case of multi-trace actions.
\subsubsection*{Abelian case}

As a warm up, consider the case of a single brane, for which the leading term in the effective action is simply the non-relativistic kinetic term\footnote{Throughout this paper, we take units in which $c=1$ and the particle mass is set to 1.}
\be
\label{kin}
S_0 = \int dt \; {1 \over 2} \dot{x}^2 \; .
\ee
This action is Galilean invariant but not Lorentz invariant. If we demand invariance under the boost transformation in (\ref{abelrules}), we must add a higher order term ${1 \over 8} \dot{x}^4$ to the action so that the variation of (\ref{kin}) under the second term in (\ref{abelrules}) is cancelled by the variation of this term under the first term in (\ref{abelrules}). The variation of the new term under the second term in (\ref{abelrules}) must be cancelled by the variation of yet a higher order term, and so forth. Of course, we know it is possible to carry this out to all orders, with one possible Lorentz invariant completion being the relativistic kinetic term
\be
\label{lorkin}
S = -\int dt \sqrt{1-\dot{x}^2} = -\int ds \; .
\ee
This result is not unique, since there are higher order Lorentz invariant structures we could add with arbitrary coefficients. The first of these is
\be
\label{nextinv}
-\int d s \( {d^2 x^\mu \over d s^2} {d^2 x_\mu \over d s^2} \) = \int dt \; \({\ddot{x}^2 \over (1-\dot{x}^2)^{3 \over 2}} + {(\dot{x}^i \ddot{x}^i)^2 \over (1- \dot{x}^2)^{5 \over 2}} \) \; ,
\ee
where $s$ is proper time, and generally, we will have one Lorentz invariant structure for each Galilean (and time-reversal) invariant leading term. On the other hand, (\ref{lorkin}) is the unique Lorentz invariant action depending on $\dot{x}$ and no higher derivatives of $x$.

We would now like to see how these statements generalize to the non-abelian case.

\subsubsection*{Constraints for leading order invariant terms}
\label{leadingorderinvariant}

Ideally, we would like to be able to write down the most general Poincar\'{e} invariant action depending on the matrix $X$ and its derivatives. Such an action would be a general linear combination of all possible independent Poincar\'{e} invariant structures with arbitrary coefficients. As a first step, we will determine a set of necessary conditions that the leading term (with the fewest $X$s) in any such structure must satisfy.

Apart from the boost transformation law, the remaining Poincar\'{e} transformations in (\ref{nonabelrules})
do not mix terms with different numbers of $X$s, so the leading term must be a rotational scalar,
have no explicit time dependence, and be invariant under a shift in $X$ by a multiple of the unit matrix,
\be
\label{cond1}
\partial_\epsilon S_0(X + \epsilon) = 0 \; .
\ee
We must also have invariance of the leading term under parity and time-reversal  transformations, and this requires an even number of $X$s and an even number of time derivatives respectively. Finally, the leading term must be invariant under the $X^0$ term in the boost transformation rule, since the variation of the full set of terms under the full transformation law will contain no other terms of this order. Thus, we must also have
\be
\label{cond2}
\partial_\beta S_0(X + \beta t) = 0 \, .
\ee
Note that these are the same conditions as in the abelian case, and are simply the statement that the leading term must be invariant under the Galilean group (including parity) plus time reversal transformations.

It is obvious that any term for which all $X$s and $\dot{X}$s appear in commutators satisfies (\ref{cond1}) and (\ref{cond2}), since in this case, even the variation of the Lagrangian is zero. More generally, we may have terms for which the variation of the Lagrangian in (\ref{cond1}) or (\ref{cond2}) is a total derivative. One example is the non-abelian generalization of (\ref{kin}),\footnote{Throughout this paper, we assume that the world-volume gauge field $A_0$ on the D0-brane world-volume has been set to zero by a gauge transformation.}
\be
\label{nonkin}
S_0 = \int dt \; \tr(\dot{X}^2) \; .
\ee
In appendix \ref{leadingproof1}, we show that this is the only example which cannot be rewritten by partial integration as a term for which all $X$s and $\dot{X}$s appear in commutators.

Thus, the lowest-order term of any Poincar\'{e} invariant action is either (\ref{nonkin}), or can be written as the integral of a scalar Lagrangian with no explicit time-dependence such that all $X$s and $\dot{X}$s appear in commutators.

In section \ref{leadingproof2}, we will argue that these necessary conditions on the leading term are actually sufficient to guarantee the existence of a Poincar\'{e} invariant completion. For now, in order to gain some confidence in this statement,
we will construct the completions order-by-order in a couple of examples using the order-by-order results from section \ref{orderbyorder} for the transformation law.

\subsubsection*{Order-by-order construction of invariant actions.}

First, we consider the simplest possible Galilean invariant potential term,
\be
\label{leadpot}
S = C \int dt \; \tr \left({1 \over 4} [X^i, X^j]^2 \right) \; ,
\ee
present in the low-energy effective action for D0-branes in weakly coupled string theory ($C$ is some dimensionful constant). In this case, the first required corrections are at ${\cal O}(X^6)$ and take the form
\bea
\label{potcor}
S &=& C \int dt \ \STr \left(
 {1 \over 4} [X^i, X^j]^2   + \right. \\
&& \left.\qquad + {1 \over 2} [X^i,X^k][X^j,X^k]\dot{X}^i \dot{X}^j - {1 \over 8} [X^i,X^j][X^i,X^j]\dot{X}^k \dot{X}^k \right) + {\cal O}(X^8), \nn
\eea
independent of the choice for the ${\cal O}(X^4)$ and higher order terms in the transformation law.
It turns out that these correction terms reproduce known terms in the D0-brane effective action obtained \cite{tseytlin}
by T-dualizing the simplest (symmetrized) non-abelian generalization of the Born-Infeld action for D9-branes.
Indeed, it may be checked that the correction terms in (\ref{potcor}) are precisely the ${\cal O}(X^4 \dot{X}^2)$ terms
in
\be
\label{tseytlin}
S = -\int dt \ \STr\left(\left[\det(\delta_{ij} + F_{ij})(1-\dot{X}^i(\delta + F)^{-1}_{ij}\dot{X}^j \right]^{1 \over 2}\right) \; ,
\ee
where $F_{ij} \equiv {\alpha'}^{-1} i [X^i, X^j]$. On the other hand, the required correction to (\ref{potcor})
at order $X^8$ includes terms with $\dot{X}$ in commutators which are not reproduced by (\ref{tseytlin}).

As a second example, we consider the simplest possible leading term, the non-relativistic kinetic term (\ref{nonkin}).
Using the boost transformation rule (\ref{nonabelrules},\ref{firstnon}) up to order $X^4$,
we find that adding the symmetrized version of terms in the abelian relativistic kinetic term suffices up to ${\cal O}(X^5)$ to make the action invariant,
but this breaks down at ${\cal O} (X^6)$.
Fortunately, it is possible to add terms involving commutators at this order to restore Poincar\'{e} invariance.
For example, using (\ref{nonabelrules},\ref{firstnon}) for the boost transformation rule, we find that the variation of
\bea
\label{exampact}
S &=& -\int dt\ \STr \left( 1 - \frac{1}{2}\dot{X}^2 - \frac{1}{8}(\dot{X}^2)^2 - \frac{1}{16} (\dot{X}^2)^3 - \right. \\
 & & \qquad \qquad - \frac{1}{24}\left(\dot{X}^i\dot{X}^k[\dot{X}^i,\dot{X}^j][\dot{X}^k,\dot{X}^j]- 3\ddot{X}^i\dot{X}^j\dot{X}^k[\dot{X}^k,[X^j,\dot{X}^i]] -\right. \nonumber\\
 & & \qquad \qquad  \left. \left. -3\ddot{X}^i\dot{X}^k[\ddot{X}^i,\dot{X}^j][X^k,X^j]   + 2\ddot{X}^j\dot{X}^k[X^k,\dot{X}^i][\dot{X}^j,\dot{X}^i]\right)\right)+ {\cal O}(X^8). \nn
\eea
is zero up to ${\cal O}(X^7)$ terms that would presumably be cancelled by the leading order variation of ${\cal O}(X^8)$ corrections to the action.
The corrections here are not among the known terms appearing in (\ref{tseytlin}).  We will see in the next subsection that these commutator correction terms can actually be eliminated by a field redefinition.

The expressions in this section are certainly not unique,
since we can always add with arbitrary coefficients any of the higher order invariant structures discussed in the previous subsection.
However, the absence of any obstruction to our order-by order construction at the first non-trivial order can be taken as evidence that a full Poincar\'{e} invariant completion exists.
In section \ref{leadingproof2}, we will provide stronger evidence and suggest a way to write manifestly invariant actions in terms of new covariant objects.

\subsubsection*{Non-abelian generalization of the relativistic kinetic term}

In discussing the abelian case, we noted that among all invariant actions, there is a special choice, the relativistic kinetic term (\ref{lorkin}), which depends only on $\dot{x}$ and not on any higher derivatives. To close this section, we would now like to see to what extend this generalizes to the non-abelian case.\footnote{This section is not essential to the development in the remainder of the paper. The reader only interested in the result may skip to the final summary paragraph on a first reading.}

To start, we show that any Poincar\'{e} invariant structure depending only on $\dot{X}$ must (apart from additive and multiplicative constants) begin with the term (\ref{nonkin}). For, assume the Lagrangian for some other invariant action $S(\dot{X})$ had a different leading term $L_n$ of order $X^n$. According to the constraints of the previous subsection, $L_n$ must have all $\dot{X}$s in commutators so that the condition (\ref{cond2}) holds, and will necessarily have $n \ge 4$.
The leading contribution to the variation of this term comes from the second term of the boost transformation in (\ref{nonabelrules}),
and using the cyclicity of the trace, we can write
\[
\delta_\beta^2 L_n = \tr(\sym(\beta^j X^j \ddot{X}^i + \beta^j \dot{X}^j \dot{X}^i) {\cal C}_{n-1}^i(\dot{X})) \; .
\]
If the full action is invariant, this variation must combine with the variation of a higher order term under the first term in (\ref{nonabelrules}) to give a total derivative
\[
\delta_\beta^2 L_n + \delta^0_\beta L_{n+2} = {d \over dt} \tr(\beta^j X^j Q_n(\dot{X})) \; .
\]
Note that we cannot have terms where $\beta$ is contracted with a derivative of $X$ on the right side since this would produce $\beta^j \ddot{X}^j$ terms which are not present on the left side. Comparing all terms containing a second derivative of $X$, we have
\be
\label{impos}
\tr(\sym(\beta^j X^j \ddot{X}^i ) {\cal C}_{n-1}^i(\dot{X})) =  \tr(\beta^j X^j {d \over dt} Q_n(\dot{X})) \; .
\ee
Now, on the left side, the $\beta^j X^j$ always appears adjacent to the $\ddot{X}$ in the trace. On the right side, $Q$ is of order $X^{n \ge 4}$, so there will certainly be terms for which the $\ddot{X}$ is not adjacent to $\beta^j X^j$. Thus, (\ref{impos}) is impossible, and our assumption that there exists a Poincar\'{e} invariant action depending only on $\dot{X}$ whose leading term is not (\ref{nonkin}) must be false.

It follows immediately that given the non-abelian transformation rules, there can be at most one independent invariant action depending only on $\dot{X}$. If there were more than one, then at least one linear combination would have a leading term other than (\ref{nonkin}), and we have seen that this is impossible.

The present result is not quite as strong as it may sound. Since we have assumed a specific transformation law, what we have actually shown is that for any given choice of field, there is at most one action depending only on $\dot{X}$. On the other hand, there could be other independent actions which after appropriate field redefinitions depend only on $\dot{X}$. In the absence of some canonical choice for the field there would be no sense in which one of these actions would be preferred over another and therefore no canonical generalization of (\ref{lorkin}) to the non-abelian case.
Actually, we will now see that any Poincar\'{e} invariant generalization of (\ref{lorkin}) to the non-abelian case can be brought to a form which depends only on $\dot{X}$, using a suitable field redefinition. In fact, for any invariant kinetic action, there is a choice of field for which the action takes the form
\be
\label{symsqrt}
S = - \int dt \; \STr \left( \sqrt{1 - {\dot{X}}^2} \right) \; .
\ee

For, consider the most general Poincar\'{e} invariant action of the form
\be
\label{general}
S = \int dt \; \tr\({1 \over 2} \dot{X}^2 + \cdots\) \; .
\ee
We assume that all higher-order terms have the same number of $X$s as time derivatives, since the variation of any other terms will not mix with the variation of these terms under Poincar\'{e}-transformations. Now, consider the lowest order terms with second or higher derivatives of $X$, or with $\dot{X}$ appearing in a commutator. These terms must be translation invariant, so may be written with all undifferentiated $X$s appearing in commutators. The terms involving higher derivatives may clearly be written as
\be
\label{lowder}
\int dt \; \tr(\ddot{X}^i F^i(X)) \; ,
\ee
for some $F$, where we can use integration by parts to put any terms with three or more derivatives on $X$ in this form. Terms with no higher derivatives but some $\dot{X}$ appearing in a commutator will be functions of $\dot{X}$ alone, so integrating by parts to remove the derivative from some $X$ appearing in a commutator will leave a set of terms all of which have a single $\ddot{X}$. Rearranging commutators in some terms, we may again bring this set of terms to the form (\ref{lowder}).

In both cases, the resulting $F$ will still have all undifferentiated $F$s appearing in commutators.\footnote{Terms involving only $\dot{X}$ which do not contain any commutators may also be brought to the form (\ref{lowder}), but in this case, $F$ will not be translation invariant.} Also, since the total number of time derivatives and $X$s was assumed to be equal, $F$ will contain at least one undifferentiated $X$, which must therefore appear in a commutator, so $F$ vanishes in the abelian case. Thus, $F$ satisfies all of the conditions listed below (\ref{FR}) for an allowed field redefinition
\[
X^i \to X^i + F^i(X) \; .
\]
Under such a field redefinition, the leading modification to the action will come from the change of the leading term in (\ref{general}) and give (after integrating by parts)
\[
S \to S - \int dt \; \tr(\ddot{X}^i F^i(X)) + {\rm higher \; \; orders} \; .
\]
which eliminates the lowest order terms in $S$ with either higher derivatives or $\dot{X}$ appearing in a commutator. By repeated field redefinitions, we can achieve this at any order, ending up with an action that contains no higher derivative terms and no commutators (i.e. a completely symmetrized function of $\dot{X}$). All terms in such an action survive in the abelian case, for which the unique Poincar\'{e} invariant function of $\dot{x}$ is (\ref{lorkin}), so our resulting action must be precisely (\ref{symsqrt}). At this point, we have fixed the choice of field completely, since any further field redefinitions will introduce additional terms into the action.

To summarize the results of this section, we have shown first that for a given definition of the field, there is at most one Poincar\'{e} invariant action depending on $\dot{X}$ and no higher derivatives. On the other hand, we have shown any invariant generalization of (\ref{lorkin}) may be written in this way by an appropriate field redefinition, and there will be a unique choice of field for which this action takes the form (\ref{symsqrt}). Thus, among the many invariant non-abelian generalizations we expect for the relativistic kinetic term with a particular choice of transformation law, there is no obvious way to make a canonical choice.

\section{Covariant objects}

The naive order-by-order approach to writing down Poincar\'{e} invariant actions discussed in the previous section
is cumbersome to say the least. This motivates us to search for a set of covariant objects,
which transform simply under Poincar\'{e} transformations, to serve as the basic building blocks
for constructing manifestly invariant actions (just as we employ the field strength in non-abelian gauge theories or the Riemann tensor in gravitational theories).

Some hope for the success of this approach may be gained from our previous study \cite{old} (see also \cite{reallyold}) of how to implement invariance under spatial diffeomorphisms for D0-branes in curved space.
There, the transformation rule for matrices $X^i$ under spatial diffeomorphisms was also extremely complicated,
but we proved (assuming the existence of a consistent transformation rule) the existence of a covariant object $V^i(y)$ built from $X$ and the metric,
transforming as a vector field under spatial diffeomorphisms. In the abelian case, this object reduces to the vector field (well-defined in some neighbourhood of the brane)
which at any point $y$ points in the direction of the geodesic from $y$ to $x$ with length equal to the geodesic distance.\footnote{Alternatively, this is the field whose exponential map gives the constant $x^i$, proportional to the spatial derivative of the geodesic distance to the brane.} In terms of $V$, we were able to write the most general invariant action as an integral over space-time
\be
\label{oldact}
\int d^{d} y \sqrt{-g(y)}\tr({\cal L}(V(y), g(y), R_{ijkl}(y), \cdots) \delta(V(y))) \; ,
\ee
where ${\cal L}$ is some scalar Lagrangian density built from $V$, the metric, and covariant derivatives of the Riemann tensor.
The object $\delta(V)$ (see \cite{old} for the precise definition in the non-abelian case) generalizes $\delta^d(x^i-y^i)$ and localizes the action to the brane locations in the case of diagonal $X^i$ where the branes have well-defined positions.

Based on this success, it is plausible that a similar construction may allow us to write actions which are manifestly invariant under general diffeomorphisms, and in particular, under the Poincar\'{e} transformations that we study in this work. Thus, we begin by searching for an appropriate generalization of the vector field $V^i(y)$.

\subsubsection*{A covariant vector field}

For the case of a single D0-brane, the spatial vector field $V^i(y)$ above has a very natural generalization to a space-time vector field $v^\mu(y)$
which contains the same information as the static gauge embedding coordinates of the brane $x^\mu(t)=(t,x^i(t))$.
Through any point $y$ sufficiently close to the brane,
there is a unique geodesic that intersects the brane world-line orthogonally (with respect to the Lorentzian metric),
and we define $v^\mu(y)$ to be the vector along this geodesic for which $v^\mu v_\mu$ is the squared geodesic distance.\footnote{Note that $v$ will not be well-defined globally unless the trajectory is non-accelerating. For a uniformly accelerating trajectory, $v$ will cease to be well-defined beyond the Rindler horizon.}

In flat space (which we restrict to in this paper) $v^\mu(y)$ is the displacement vector from $y^\mu$ to the point $x^\mu(t_y)$ that is simultaneous with $y^\mu$ in the instantaneous rest frame of the brane. In other words,
\be
\label{vdef}
v^\mu(y)\equiv x^\mu(t_y)-y^\mu,
\ee
where $t_y$ is implicitly determined by the condition
\be
\label{tdef}
v_\mu(y) \dot{x}^\mu(t_y)=0.
\ee
For an accelerating brane, planes orthogonal to the brane will generally intersect each other at points sufficiently far away,
so it is clear that $v^\mu(y)$ is not globally well-defined. However, this is enough to ensure that there is a well-defined expansion for $v^\mu(y)$ in powers
of the static gauge coordinates $x^i(t)$ and its derivatives, and it is this expansion that we will use primarily in what follows.

Since our definition of $v^\mu(y)$ was coordinate-independent, this must transform as a four-vector under Lorentz transformations,
\[
\tilde{v}^\mu (\Lambda y) = \Lambda^\mu{}_\nu v^\nu(y) \; .
\]
In particular, for an infinitesimal boost we have
\bea
\delta v^0(t,\v{y}) & = & \v{\b}\cdot \v{v}\; (t,\v{y}) - t \v{\b}\cdot \v{\nabla} v^0(t,\v{y}) - \v{\b}\cdot \v{y}\; \d_t v^0(t,\v{y}), \nn \\
\delta \v{v}\; (t,\v{y}) & = & \v{\b}\; v^0(t,\v{y}) - t \v{\b}\cdot \vec{\nabla} \v{v} (t,\v{y}) - \v{\b}\cdot\v{y}\; \d_t \v{v}\; (t,\v{y}).
\label{covvect}
\eea

\subsubsection*{Generalization to the case of multiple D0-branes}

We would now like to see whether $v^\mu$ generalizes to the non-abelian case. That is, we would like to construct a set of matrix-valued functions $V^\mu(y)$ defined as a formal expansion in terms of $X^i(t)$ which transform as a space-time vector field and which reduce to ${\rm diag}(v_{x_1}^\mu(y),\dots, v_{x_N}^\mu(y))$ when the matrices $X^i(t)$ are diagonal. To ensure the latter condition, we may write
\be
\label{nongen}
V^\mu(y)= V^\mu_{\rm sym}(y) + \Delta V^\mu(y),
\ee
where $V^\mu_{\rm sym}(y)$ is the expression obtained by replacing all occurrences of $x^i$
in the expansion of $v^\mu(y)$ with $X^i$ and using the completely symmetrized product of matrices and $\Delta V^\mu(y)$ is an expression that must involve commutators.

To see whether the construction is possible, we write the most general expansion of the form (\ref{nongen}) up to some order in $X$, and demand that the covariant transformation rules (\ref{covvect}) are satisfied to this order using the order-by-order results for the transformation rule obtained in section 2. Happily, we find that at least up to order $X^5$, it is possible to choose $\Delta V^\mu(y)$ so that the covariant transformation rules hold.
The success of this procedure is already quite nontrivial at order $X^4$ for which we give the results in appendix \ref{navectorfield}.

Unfortunately, we do not have a proof that an appropriate $V^\mu(y)$ can be constructed to all orders.
If it can, it is easy to see that many such objects exist,
since we may always construct others from the original one e.g.\ $\tilde{V}^\mu = V^\mu + \partial_\rho V_\nu [V^\mu, \partial^\nu V^\rho]$.
There may be some canonical choice for $V^\mu$, as we found in \cite{old},
but we do not know the additional constraints that would select this.\footnote{One constraint that we might impose is that $V$ should satisfy
$\partial_\mu V_\nu = \partial_\nu V_\mu$.
This holds in the abelian case, since $V_\mu = -{1 \over 2} \partial_\mu V^2$.
In the non-abelian case, given any definition of $V^\mu$ we can take $\tilde{V}^\mu = -{1 \over 2} \partial^\mu (V^\nu V_\nu)$ which ensures
that $\tilde{V}^\mu$ is covariant and that $\partial_\mu \tilde{V}_\nu$ is symmetric.} On the other hand, we will see that any choice for $V$ (assuming one exists) will allow us to construct the most general Poincar\'{e} invariant actions.

\subsubsection*{A covariant matrix distribution}
\label{mdist}

Assuming that the covariant object $V^\mu(y)$ exists in the non-abelian case, it is now trivial to construct scalar fields ${\cal L}(y)$ simply by taking any product involving $V^\mu$ and its derivatives such that all indices are contracted with $\eta^{\mu \nu}$. To obtain an invariant action, we should integrate over space-time, but we still need some analogue of the $\delta(V(y))$ term in (\ref{oldact}) that would localize the action to well-defined world-volumes of the individual branes in the case of diagonal $X^i$. We have not been able to construct such a distribution directly from the covariant object $V^\mu$. However, we find in this subsection that it is possible to construct an object with the appropriate transformation properties directly, at least up to two commutator terms to all orders in $X$.

Our goal is to construct from $X^i(t)$ a matrix valued field $\Theta(y)$ such that actions of the form
\be
\label{action0}
S=\int d^{d+1}y\ \Tr \big( {\cal L}(y) \Theta(y) \big),
\ee
will be invariant if ${\cal L}$ is a scalar built from $V^\mu$. Here, $\Theta(y)$ should transform as a density and should contain the matrix generalization of a
delta function reducing the integral over $d+1$-dimensional space-time
to an integral over the one-dimensional world-sheet. In other words, it is the matrix generalization of the distribution $\theta(y)$ for the single brane
case, which takes the form
\be
\theta(y)= \int d\tau \sqrt{- \d_\tau x^\mu \d_\tau x_\mu} \; \delta^{d+1}(x^\nu(\tau)-y^\nu) \; .
\label{abeliandistr}
\ee
We will call it the {\em covariant matrix distribution}.

Under Lorentz transformation a density should transform as
\be
\tilde{\Theta}(\Lambda y) = \Theta(y),
\ee
or specifically under an infinitesimal boost
\be
\label{boostcon}
\delta_\b \Theta(t,\vec{y}) = - t \vec{\beta} \cdot \vec{\nabla} \Theta(t,\vec{y}) - \vec{\beta} \cdot \vec{y} \; \partial_t \Theta(t,\vec{y}).
\ee
Defining the moments of the distribution as
\be
\Theta^{(i_1 \cdots i_n)}(t) = \int d^dy \  \Theta(t,\vec{y})\; y^{i_1} \cdots y^{i_n} \; ,
\label{moments}
\ee
we find that the constraints (\ref{boostcon}) become
\be
\delta \Theta^{(i_1\ldots i_n)}=n t \b^{(i_1} \Theta^{i_2\ldots i_n)} - \b_l \frac{d}{dt}\Theta^{(l i_1 \ldots i_n)}.
\label{constraintdistr}
\ee
To zeroth order in the commutators a solution to this constraint is given by
\be
\Theta_{\rm sym}^{(i_1 \ldots i_n)}=\sym \(\sqrt{1-\dot{X}^2}\; X^{(i_1} \ldots X^{i_n)}\).
\label{distlowest}
\ee
In fact, this is the only solution (modulo an overall constant and
rescaling of $X$) build solely out of $X$ and $\dot{X}$.  In terms of
the density we have at leading order
\be
\Theta_{\rm sym}(t,\vec{y})=\sym \(\sqrt{1-\dot{X}^2}\; \delta^d(X(t)-y)\) ,
\label{distlowest2}
\ee
where
\[
\delta^d(X-y) \equiv \int {d^dk \over (2\pi)^d} \; e^{i k^i (X-y)^i} \; ,
\]
so that it indeed contains the required $d$-dimensional delta-functions in the case where $X$ is diagonal.

We must now ask whether it is possible to add correction terms to (\ref{distlowest}) such that the constraints (\ref{constraintdistr}) are satisfied with the non-abelian transformation rules (\ref{nonabelrules}, \ref{2commtranslawterm}). While we have not been able to prove this to all orders, we have checked through a lengthy calculation that a solution exists up to second order in commutators but to all orders in $X$.

The two-commutator calculation is described in appendix \ref{2commdistrcalc}. We found that the calculation simplified using the ansatz
\be
\Theta^{(i_1\ldots i_n)}=\sym \(\sum_p C_n^{'(i_1 \ldots i_p} X^{i_{p+1}} \ldots X^{i_n)}\),
\ee
with
\bea
C_n^{'(i_1\ldots i_n)} & = & \sqrt{1-\dot{X}^2} \; C_n^{(i_1\ldots i_n)} + \frac{\d}{\d \dot{X}_{j}}\sqrt{1-\dot{X}^2} \, E_n^{j;(i_1 \ldots i_n)} \nn \\
& & + \sum_k \frac{\d}{\d \dot{X}_{j_1}}\ldots \frac{\d}{\d \dot{X}_{j_k}}\sqrt{1-\dot{X}^2} \, R^{(j_1 \ldots j_k);(i_1\ldots i_n)} \; .
\label{sqrtansatz}
\eea
Explicit results for the tensors $C$, $E$, and $R$ appear in appendix \ref{2commresults}. Up to two commutators we can take $C_p=0$ for $p\ge 4$ and $E_p=0$ for $p \ge 3$. Also, we can take all $R$ tensors zero except for $R^{(j_1j_2);i_1}$, $R^{(j_1j_2)}$ and $R^{(j_1j_2j_3)}$.

The result given in the appendix is not unique, but we will see below that any specific choice for $\Theta$ and $V$ will be enough to generate all possible Lorentz invariant actions. From now on, we will assume that covariant $V^\mu(y)$ and $\Theta(y)$ exist to all orders, and proceed to discuss the Poincar\'{e} invariant actions.

\section{Manifestly Lorentz invariant D0-brane actions}
\label{maninvact}

Given the vector field $V^\mu(y)$ and the covariant matrix distribution $\Theta(y)$, it is now manifest that any action
\be
\label{newact}
\int d^{d+1} y \; \tr({\cal L}(y) \Theta(y))
\ee
will be invariant as long as ${\cal L}$ is a scalar field built from $V$ and its derivatives.
To obtain an explicit expansion of this action in powers of $X$ and its time derivatives, we may use the expansion
\be
\Theta(t,\v{y}) = \sum_{p=0}^\infty \frac{(-1)^p}{p!} \Theta^{(i_1\dots i_p)}(t)\ \d_{i_1}\dots \d_{i_p} \delta^d(y),
\ee
of $\Theta$ in terms of its moments.  Then the action takes the form
\bea
S &=&\int dt\ \sum_{p=0}^\infty \Tr\left(\d_{i_1}\dots \d_{i_p}{\cal L}(V)|_{y^i=0}\ \Theta^{i_1\dots i_p}(t) \right).
\eea
Since $\Theta^{(i_1\dots i_p)}={\cal O}(X^p)$, the leading term in the action will come from the set of all terms
for which $n + {\it order}(\partial_{i_1} \cdots \partial_{i_n} {\cal L})$ is a minimum.

The expression (\ref{newact}) clearly gives rise to a large class of invariant actions. It turns out that any invariant action can be written in this way, as we now show.

\subsubsection*{The most general Poincar\'{e} invariant action}
\label{leadingproof2}

In section \ref{leadingorderinvariant}, we showed that the leading term $S_0$ of any Poincar\'{e} invariant action could be written using a rotational scalar Lagrangian built from an even number of $X$s and an even number of time-derivatives,
such that $S_0 = \int dt \; \tr(\dot{X}^2)$ or all $X$s and $\dot{X}$s appear in commutators. We will now show that any term satisfying these conditions has a Poincar\'{e} invariant completion that may be written in the form (\ref{newact}), and that these completions form a basis for the full set of Poincar\'{e} invariant actions.

First, if $S_0 = \int dt \; \tr(\dot{X}^2)$, we can write a Poincar\'{e} invariant completion as
\[
-\int d^{d+1} y \; \tr(\Theta(y)) \; .
\]
Otherwise, the leading order Lagrangian $L_0$ may be written as a sum of terms for which all $X$s and $\dot{X}$s appear in commutators. Now $L_0 = \tr({\cal L})$ is a rotational scalar, and by parity and time reversal invariance, must have an even number of $X$s and an even number of $\dot{X}$s. Consequently, the index on each matrix $(X^i)^{(m)}$ will pair with the index on some other matrix $(X^i)^{(n)}$ where $m$ and $n$ are the number of time derivatives on the first and second matrix respectively.\footnote{In particular, any terms involving an odd number of $\epsilon$ tensors will not be invariant under parity/reflections, while terms involving an even number may be rewritten using $\delta$s. There will be additional structures involving $\epsilon$s which are invariant under the part of the Poincar\'{e} group continuously connected to the identity but violate either parity or time translation invariance; we will not discuss them further here.} We now define a matrix object ${\cal L}(y)$ built out of $V^\mu$ by making the following replacements in ${\cal L}$, depending on whether $m$ and $n$ are both even, both odd, or of opposite parity. If $m$ and $n$ have the same parity, we replace
\bea
\label{rep1}
(X^i)^{(2k)} \cdots (X^i)^{(2l)} \qquad &\rightarrow& \qquad \partial^{(2k)} V^\mu \cdots \partial^{(2l)} V_\mu \; , \cr
(X^i)^{(2k+1)} \cdots (X^i)^{(2l+1)} \qquad &\rightarrow& \qquad -{1 \over 2} \partial^{(2k)} \partial_\mu V^\nu \cdots \partial^{(2l)} \partial_\nu V^\mu \; .
\eea
Since the total number of time derivatives is even, there must be an even number of pairs where $m$ and $n$ have opposite parity. We may then group these arbitrarily into pairs of paired $X$s, and make the replacement
\bea
(X^i)^{(2k)} \! \cdot \! \cdot \! \cdot \! (X^i)^{(2l+1)} \! \cdot \! \cdot \! \cdot \! (X^j)^{(2p)} \! \cdot \! \cdot \! \cdot \! (X^j)^{(2q+1)}  \rightarrow  -\partial^{2k} V^\mu \! \cdot \! \cdot \! \cdot \! \partial_\mu \partial^{2l} V^\alpha \! \cdot \! \cdot \! \cdot \! \partial^{2p} V^\nu \! \cdot \! \cdot \! \cdot \! \partial_\nu \partial^{2q}   V_\alpha . \cr
\label{rep2}
\eea
After these replacements, we are left with an object ${\cal L}$ that transforms as a scalar field, so the action
\be
\label{compl}
\int d^{d+1} y \; \tr({\cal L}(y) \Theta(y))
\ee
will be Poincar\'{e} invariant. Furthermore, it is easy to check that in the replacements (\ref{rep1}) and (\ref{rep2}),
the contributions on the right side which are of lowest order in $X$ have $y$-independent terms which are precisely the terms on the left.
It is important here that all expressions $V^\mu$ and $\partial_\nu V^\mu$ appear in commutators
(since we assume all $X$s and $\dot{X}$s do), so that possible lower order terms from the leading $y^i$ in $V^i$ vanish.
As a result, the leading order term in ${\cal L}(y=0)$ is exactly ${\cal L}$,
and all of the $y$-dependent terms in ${\cal L}(y)$ will only lead to higher order terms in the action,
so the action (\ref{compl}) will have leading term $\int dt \; L_0$. This completes the proof (assuming the existence of covariant objects ${\cal L}$ and $\Theta$) that all Galilean and time-reversal invariant leading terms have Poincar\'{e} invariant completions that can be written in the form (\ref{newact}).

To show that the terms just constructed form a basis for all Poincar\'{e} invariant actions, let us suppose this were not true. Then consider some action $S$ linearly independent from the set ${S_i}$ we have just constructed. Then among all actions $S-\sum c_i S_i$ there must a subset whose leading terms have maximum order. Choose an action $S_{max}$ in this subset, and suppose that $S_{max}$ has leading term $S_0$ at order $X^p$. By the results in section 2, this term must be Galilean and time-reversal invariant, and we have just seen that $S_0$ has some Poincar\'{e} invariant completion $S'$ that can be written in the form (\ref{newact}). But then $S_{max} - S'$ is of the form $S-\sum c_i S_i$ and has a leading term of higher order than $S_{max}$, contradicting our assumption.

To summarize, we have now shown that every Poincar\'{e} invariant action has a Galilean and time-reversal invariant leading term, and any such term has a Poincar\'{e} invariant completion that may be written in the form (\ref{newact}).
Finally, the set of such terms form a basis for all possible Poincar\'{e} invariant actions.

\subsubsection*{Examples}
\label{applications}

To close this section, we discuss as examples the Poincar\'{e} invariant completions of the simplest kinetic and potential terms.

First, by the results of this section, the most general Poincar\'{e} invariant completion of the kinetic term (\ref{nonkin}), allowing only terms with as many time derivatives as $X$s (i.e.\ terms that can mix with the leading term under a Lorentz transformation) is\footnote{Note that any choice for ${\cal L}_4$ may be absorbed into a redefinition of $\Theta(y)$. The arbitrariness in $\Theta$ corresponds to the freedom to make such redefinitions.}
\[
\int d^{d+1} y \; \tr(\Theta(y)(-1 + {\cal L}_4(V(y)))) \; ,
\]
where ${\cal L}_4$ is an arbitrary scalar built from $V$s and an equal number of derivatives, which may without loss of generality be taken to be a term with at least two commutators of order $V^4$ or higher.\footnote{This follows since the leading term in any higher order invariant action will be at least of order $X^4$ and by the construction of the previous subsection, we may construct such an action using an ${\cal L}$ with terms of order $V^4$ and higher.}
While the result is by no means unique, it is highly constrained relative to the set of all possible translation and rotation invariant actions.

As a precise example of the degree to which the action has been constrained, consider all terms with up to two commutators. In this case, there are only a finite number of independent terms in ${\cal L}_4$ that can contribute. To see this, note that the leading term of any such expression may be written schematically as
\[
\STr([X,X][X,X] X \cdots X) \; ,
\]
where the total number of $X$s is $4 + 2n$ for some $n$, and the total number of time derivatives must be equal to this. For Galilean invariance, all $X$s outside commutators must have at least two time derivatives, so there must be at least $4n$ time derivatives. Then
\[
4n \le 2n + 4,
\]
so we have $n \le 2$. It is then easy to write down all possible leading terms containing two commutators; up to total derivatives we find 8, 17, and 2 terms respectively for $n$ equal to 0, 1, and 2. Thus, the most general Poincar\'{e} invariant completion of the kinetic term (\ref{nonkin}) contains 27 arbitrary coefficients up to terms involving more than two commutators. On the other hand, the number of independent translation and rotation invariant terms with equal numbers of $X$s and time derivatives and up to two commutators is infinite, so we see that the additional requirement of boost invariance is indeed a severe constraint on the action.

Using our results for $\Theta$ and $V$, we can write explicitly the most general Poincar\'{e} invariant kinetic term up to two commutators as
\bea
S & = & -\int dt \, \STr \Bigg[ \sqrt{1-\dot{X}^2} +
\frac{\d}{\d \dot{X}^i}\sqrt{1-\dot{X}^2} \, \dot{C}_1^{i}
+ \frac{\d}{\d \dot{X}^i}\frac{\d}{\d \dot{X}^j}\sqrt{1-\dot{X}^2} \, R_0^{(ij)} \nn \\
& & + \frac{\d}{\d \dot{X}^i}\frac{\d}{\d \dot{X}^j}\frac{\d}{\d \dot{X}^k}\sqrt{1-\dot{X}^2} \, R_0^{(ijk)} \Bigg]
+ \int d^{d+1} y \tr(\Theta_{\rm sym}(y) {\cal L}_4(V_{\rm sym}(y)))  +  \mathcal{O}([\cdot,\cdot]^3), \nonumber \\
\eea
where $C_1^i$, $R_0^{(ij)}$ and $R_0^{(ijk)}$ are defined in appendix \ref{2commresults}.
Here $\Theta_{\rm sym}$ and $V_{\rm sym}$ are the symmetrized generalization of the abelian expressions for $\Theta$ and $V$, while ${\cal L}_4$
may be obtained by promoting the general linear combination of our 27 Galilean invariant two-commutator terms to Lorentz-scalar expressions built from $V$.
Up to order $X^6$ this reduces to our earlier result (\ref{exampact}) plus the general linear combination of the 25 independent Galilean invariant terms
with four and six time derivatives.

As a second example, we consider the Poincar\'{e} invariant completion of the potential term (\ref{leadpot}). Allowing only terms that can mix with the leading term under a Lorentz transformation, the most general invariant completion is
\[
C \int d^{d+1}y \; \tr( \Theta(y) ([V_\mu, V_\nu] [V^\mu, V^\nu] + {\cal L}_6(V(y)) )  \; .
\]
where ${\cal L}_6$ is the general linear combination of all scalars built from $n$ $V$s and $n-4$ derivatives. Without loss of generality, $n$ may be taken to be at least 6, and all terms in ${\cal L}_6$ may be taken to have at least 3 commutators. Thus, the full set of two-commutator terms in the Poincar\'{e} invariant completion of (\ref{leadpot}) are uniquely determined to be
\[
C \int d^{d+1}y \; \tr( \Theta_{\rm sym}(y) ([{V_{\rm sym}}_\mu, {V_{\rm sym}}_\nu ] [{V_{\rm sym}}^\mu, {V_{\rm sym}}^\nu]))  \; .
\]
where $\Theta_{\rm sym}$ and $V_{\rm sym}$ are the symmetrized parts of $\Theta$ and $V$. Using the explicit results for $V_{\rm sym}$ in appendix \ref{navectorfield} and the expression for $\Theta_{\rm sym}$ in (\ref{distlowest2}), we find that the full set of two-commutator terms are precisely the ones appearing in (\ref{tseytlin}).\footnote{Because of the delta function appearing in $\Theta_{\rm sym}$, all terms in $V^\mu$ of order $(X-y)^2$ and higher vanish.} The derivation of that action relied on T-duality arguments specific to string theory, so it is interesting that at least the two-commutator subset of terms can be derived purely based on Poincar\'{e} invariance.

Note that based on rotation and translation invariance alone, the full set of allowed two-commutator correction terms to potential (\ref{leadpot}) is
\beas
& & \sum_n \; \STr( b_n [X^i, X^j]^2 \dot{X}^{2n} + c_n [X^i, X^j][X^i,X^k] \dot{X}^j \dot{X}^k \dot{X}^{2n} ) \; ,
\eeas
so in this case, the additional constraint of boost invariance fixes the infinite series of coefficients $b_n$ and $c_n$ completely.

To close this section, we note that our structures $V$ and $\Theta$ provide an alternate way to write invariant actions even in the abelian case. For example, using our prescription, the Galilean invariant term ${1 \over 2} \ddot{x}^2$ has Lorentz invariant completion
\[
\int d^{d+1}y\ \partial^2 v^\mu \partial^2 v_\mu\; \theta(y) \; .
\]
Using the abelian expression (\ref{abeliandistr}) for $\theta$ and those in appendix C for $v$, this reduces precisely to the right side of (\ref{nextinv}).

\section{Lorentz covariant currents}
\label{conscurrents}

We have seen that the requirement of Poincar\'{e} invariance places severe constraints on the form of the effective action. In this section, we note that similar constraints arise in the expressions for the conserved space-time currents associated with the branes. We use the example of the D0-brane current for D0-branes in uncompactified type IIA string theory, which couples to the Ramond-Ramond one-form field of type IIA supergravity. Identical considerations apply to the other currents, which include the stress-energy tensor, the higher brane currents, and the string current (which couples to the NS-NS two-form).

The D0-brane current $J^\mu(y)$ appears in the effective action coupled to the Ramond-Ramond one-form $C_\mu$ as
\be
S = \mu \int d^{10} y \; C_\mu (y) J^\mu (y) \; .
\label{CS}
\ee
Since $C_\mu$ is a Lorentz vector, $J^\mu(y)$ must be some expression built from $X^i(t)$ transforming as a vector under Lorentz transformations. At low energies / small velocities, the leading order expression for $J^\mu(y)=(\rho(y), J^i(y))$ (ignoring fermions) is a simple generalization of the abelian expression \cite{tv0},
\bea
\label{leadcurr}
\rho (t,\v{y}) &=& \tr\left( \delta^d(X(t)-y) \right) \equiv \int {d^9 k \over (2\pi)^9} \; \tr(e^{i k^i (X-y)^i}) \; , \cr
J^i (t.\v{y}) &=& \tr\left( \dot{X}^i(t)\; \delta^d(X(t)-y) \right) \equiv \int {d^9 k \over (2\pi)^9} \; \tr(\dot{X}^i e^{i k^j(X-y)^j}) \; .
\eea
It is easy to check that current conservation,
\be
\label{currentcons}
\partial_\mu J^\mu = 0 \; ,
\ee
is satisfied with these definitions. However, we will now see that $J^\mu$ does not transform as a vector under Lorentz transformations (without additional correction terms).

A Lorentz vector field $J^\mu$ should transform under Lorentz transformations as
\bea
\tilde{J}^\mu (\Lambda y) &=& \Lambda^\mu {}_\nu J^\nu(y) \; .
\eea
This implies that under an infinitesimal boost we have
\bea
\delta_\b \rho\; (t,\vec{y}) &=& \vec{\beta} \cdot \vec{J}\; (t,\vec{y}) - t \vec{\beta} \cdot \vec{\nabla}
\rho\; (t,\vec{y}) - \vec{\beta} \cdot \vec{y} \; \partial_t \rho\; (t,\vec{y}), \nn \\
\delta_\b \v{J} (t,\vec{y}) &=&  \v{\beta}\; \rho\; (t,\vec{y}) - t \vec{\beta} \cdot \vec{\nabla}
\v{J} (t,\vec{y}) - \vec{\beta} \cdot \vec{y} \; \partial_t \v{J} (t,\vec{y}).
\eea
It is convenient to define multipole moments of the current components as in (\ref{moments}). In terms of these, the constraints of Lorentz covariance read
\bea
\delta_\b \rho^{(i_1 \cdots i_n)} &=& \v{\beta}\cdot \v{J}^{(i_1 \cdots i_n)} + n t \beta^{(i_1} \rho^{i_2 \cdots i_n)} - \beta^j \frac{d}{dt} \rho^{(j i_1 \cdots i_n)}, \nn \\
\delta_\b \v{J}^{(i_1 \cdots i_n)} &=&  \v{\beta}\; \rho^{(i_1 \cdots i_n)} + n t \beta^{(i_1} \v{J}^{i_2 \cdots i_n)} - \beta^j \frac{d}{dt} \v{J}^{(j i_1 \cdots i_n)}.
\label{cons}
\eea

Using the non-abelian transformation rules (\ref{nonabelrules},\ref{2commtranslawterm}), we may now check whether these relations are satisfied for the moments that follow from the leading expressions (\ref{leadcurr}) for the currents, namely
\bea
\rho_{\rm sym}^{(i_1 \cdots i_n) } &=& \STr (X^{i_1} \cdots X^{i_n}), \nn \\
J_{\rm sym}^{i;(i_1 \cdots i_n)} &=& \STr (\dot{X}^i X^{i_1} \cdots X^{i_n}).
\label{leadcurr2}
\eea

It is easy to check that all the constraints (\ref{cons}) are
satisfied with the expressions (\ref{leadcurr2}) in the abelian case
or for diagonal matrices, but are not satisfied in general. Thus,
the full Lorentz covariant D0-brane currents must include additional
higher order terms involving matrix commutators, and these
correction terms should be heavily constrained by (\ref{cons}).

We have checked that up to two commutator terms and to all orders in $X$ there do exist corrections to the currents such that (\ref{cons}) are satisfied. The very tedious calculation is briefly outlined in appendix
\ref{2commcurrentcalc}. It turns out that the result can be written as
\bea
\rho^{(i_1\ldots i_n)} & = & \STr \left[ \sum_p \left( \begin{array}{c} n \\ p \end{array} \right) C_p^{(i_1 \ldots i_p} X^{i_{p+1}} \ldots X^{i_n)}\right], \nn \\
J^{i;(i_1 \ldots i_n)} & = & \STr \left[ \sum_p \left( \begin{array}{c} n \\ p \end{array} \right) \( \dot{X}^i C_p^{(i_1 \ldots i_p}+E_p^{i;(i_1\ldots i_p}\) X^{i_{p+1}} \ldots X^{i_n)}\right],
\label{currenttemplate}
\eea
where $C_p$ and $E_p$ (which satisfy the same constraints as the objects of the same name appearing in the covariant matrix distribution) are given in appendix D.3.

Up to two commutators we can take $C_p=0$ for $p\ge 4$ and $E_p=0$ for $p \ge 3$.
We remark here only that $C_p$ and $E_p$ do not contain any $X$s without derivatives outside of commutators.
Furthermore $C_0=1$ which gives the leading order (\ref{leadcurr2}) and the others contain 2 commutators.
By a field redefinition, it is also possible to set $E_0^i = 0$ and $C_1^i=0$, which then fixes the field redefinition ambiguity completely;
however, it is not clear whether such a choice of field is the most natural.
The result given in the appendix is not the only possible solution. There are some terms that can be added with an arbitrary coefficient.
So as we found for the action, the constraints of Lorentz covariance are not enough to completely determine
the higher order corrections to the currents.

Current conservation (\ref{currentcons}) requires that
\be
\frac{d}{dt} C_p^{(i_1\ldots i_p)} = p E_{p-1}^{(i_1;i_2\ldots i_p)},
\label{currentcons2}
\ee
which is satisfied by the expressions in appendix \ref{2commresults}. Current conservation also assures that
the Chern-Simons action (\ref{CS}) is invariant under the gauge transformation of the R-R field, $\delta C_{\mu}=\partial_{\mu} \Lambda$.

Remarkably, in the whole calculation the cyclicity property of the trace is never used so that even without the trace,
these objects transform covariantly.  While the existence of Lorentz covariant conserved currents should have been expected by the existence of a Lorentz and gauge invariant effective action coupling the brane degrees of freedom to the bulk fields, we do no not know of any reason why untraced covariant currents should exist. These provide further examples (along with $\Theta$ and $V$) of matrix valued covariant objects.

Corrections to the D0-brane current were also proposed in \cite{Hashimoto:2004fa}. It turns out that these are part of a separate Poincar\'{e} covariant structure.

\section{Discussion}

In this note, we have focused on the simplest scenario for which a symmetry transformation on the low-energy degrees of freedom of a system of multiple D-branes mixes directions along the brane world-volume (in this case, the time direction) with directions transverse to it. We hope that the observations here will be useful in understanding the constraints that invariance under general diffeomorphisms imposes on the actions for arbitrary systems of multiple branes in curved space. We note here that at least in the abelian case, the elements $V$ and $\Theta$ in our construction generalize naturally to branes of arbitrary dimension in curved space, so we are optimistic that these elements will play a role in the general story \cite{next}.

\section*{Acknowledgements}

We would particularly like to thank Washington Taylor for a number of helpful discussions at the outset of this project. We would also like to thank Raphael Bousso, Kazuyuki Furuuchi, and Indrajit Mitra for discussions. M.V.R. would like to thank the Banff International Research Station and the Perimeter Institute for hospitality while part of this work was completed. This work has been supported in part by the Natural Sciences and Engineering Research Council of Canada and by the Canada Research Chairs Programme. H.L. acknowledges the support of the CD Howe Memorial Foundation.  M.V.R. is supported by the Alfred P. Sloan Foundation.

\appendix

\section{Calculation of the 2-commutator corrections to the transformation law}
\label{2commtranslawcalc}

To calculate the 2-commutator corrections to the transformation law we first calculate the deviation from (\ref{algebrabb}) when using the
naive boost transformation law with $T^{ij}=0$. This will require us to work out a symmetrization nested within another symmetrization,
for which we can use the following formula (up to two commutators):
\bea
\lefteqn{\sym\(A_1 \ldots A_n \sym\(B_1 \ldots B_r\)\) \sym\(A_1 \ldots A_n \, B_1 \ldots B_r\)} \nn \\
& + & \frac{1}{24} \sum_{i\neq j=1}^n \sum_{k\neq l=1}^r \sym\([A_i,B_k][A_j,B_l]\,A_1\ldots \hat{A}_i\ldots \hat{A}_j\ldots A_n \,B_1 \ldots \hat{B}_k \ldots \hat{B}_l\ldots B_r\) \nn \\
& - & \frac{1}{12} \sum_{i=1}^n \sum_{k\neq l=1}^r \sym\([B_k,[A_i,B_l]]\,A_1\ldots \hat{A}_i\ldots A_n \, B_1 \ldots \hat{B}_k \ldots \hat{B}_l\ldots B_r\) \nn \\
& + & \mathcal{O}([\cdot,\cdot]^4).
\label{nestedsym}
\eea
The result is the right-hand side of (\ref{fourth}). Now however we are more ambitious than when we solved for (\ref{fourth}) in the main text
and try to find a $T^{ij}$ such that up to second order in the commutators
\bea
\lefteqn{\b^j\delta_{\tilde{\b}} T^{ij} +\tilde{\b}^j \b^k \sym \(\dot{T}^{ik} X^j+\dot{X}^i T^{jk}\) =} \nn \\
 & & \frac{1}{8}\tilde{\b}^k \b^j \left(  - [\ddot{X}^i,[X^k,X^j]] + [\dot{X}^k,[X^j,\dot{X}^i]] - [\dot{X}^j,[X^k,\dot{X}^i]]  \right) \; .
\label{Tcommorder2}
\eea
Since we are working up to two commutators and $T^{ij}$ will already contain two commutators we can put $T^{ij}=0$
in $\delta_{\tilde{\b}}$.

At this point we need to make a suitable ansatz for $T^{ij}$.
First observe that a term of the general form
\be
A^{i_1\ldots i_q}= \sym \left[ A^{i_1 \ldots i_q;j_1\ldots j_n;k_1s_1\ldots k_p s_p}
E^{k_1 s_1}\ldots E^{k_p s_p} \(1+\dot{X}^u E^{uv} \dot{X}^v\)^p E^{j_1 t_1}\dot{X}^{t_1}\ldots E^{j_n t_n} \dot{X}^{t_n} \right],
\ee
with $E^{ks}$ defined in (\ref{Einverse}),
transforms as follows:
\bea
\lefteqn{\delta_{\b} A^{i_1 \ldots i_q} =} \nn \\
& & \sym \Big[ A^{i_1 \ldots i_q;l j_2\ldots j_n;k_1s_1\ldots k_p s_p}
E^{k_1 s_1}\ldots E^{k_p s_p} \(1+\dot{X}^u E^{uv} \dot{X}^v\)^p E^{j_2 t_2}\dot{X}^{t_2}\ldots E^{j_n t_n} \dot{X}^{t_n} \nn \\
& & - \beta^l X^l \dot{A}^{i_1 \ldots i_q} - \sum_{s=1}^q \beta^l \dot{X}^{i_s} A^{i_1 \ldots i_{s-1} l i_{s+1} \ldots k_q} + (n+2p-m) \beta^l \dot{X}^l A^{i_1\ldots i_q} \nn \\
& & + \(\delta' A^{i_1 \ldots i_q;j_1\ldots j_n;k_1s_1\ldots k_p s_p} \) E^{k_1 s_1}\ldots E^{k_p s_p} \(1+\dot{X}^u E^{uv} \dot{X}^v\)^p E^{j_1 t_1}\dot{X}^{t_1}\ldots E^{j_n t_n} \dot{X}^{t_n} \Big] \nn \\
& & + \, ({\rm higher-order\,commutators})
\label{varterm}
\eea
where $m$ denotes the number of derivatives in $A^{i_1 \ldots i_q;j_1\ldots j_n;k_1s_1\ldots k_p s_p}$ and we defined
\be
\delta'_\b A^{r_1 \ldots r_s} = \delta_{\b} A^{r_1 \ldots r_s} + \b^l X^l \dot{A}^{r_1 \ldots r_s} + \sum_{t=1}^s \b^l \dot{X}^{r_t} A^{r_1 \ldots r_{t-1} l r_{t+1} \ldots r_s}.
\label{deltap}
\ee
We put
\be
T^{il}=\sym \Bigg[\frac{1}{8} E^{jt} \dot{X}^t (- [\ddot{X}^i,[X^j,X^l]] + [\dot{X}^j,[X^l,\dot{X}^i]] - [\dot{X}^l,[X^j,\dot{X}^i]]) \Bigg] \; ,
\ee
and use (\ref{varterm}) for $q=2,p=0,n=1$ and $m=2$.
The first line in (\ref{varterm}) produces the desired compensating term on the right-hand side of (\ref{Tcommorder2}), the
second line vanishes against the other terms on the left-hand side in (\ref{Tcommorder2}) if $n+1+2p-m=0$.
We need to introduce more corrections however to cancel the third line. Continuing this process and using (\ref{varterm}) again
to make suitable ansatze for the extra terms, eventually the terms in the third line will vanish so that we do not need to add further terms and
the process ends.

\section{Characterization of Galilean invariant non-abelian actions}
\label{leadingproof1}

Here we prove that the minimal conditions (\ref{cond1}, \ref{cond2}) for the leading term
of a Poincar\'{e} invariant action imply that the leading term of the Lagrangian is either $\tr(\dot{X}^2)$
or can be written in a way such that all $X$s and $\dot{X}$s appear inside commutators.

We first show that any action satisfying
\be
\label{tran}
\partial_\epsilon S (X+ \epsilon) = 0
\ee
can be written as the integral of a Lagrangian with $\partial_\epsilon L(X+ \epsilon) = 0$ i.e.\ such that all $X$s in $L$ appear in commutators.

For suppose that an action $S$ satisfying (\ref{tran}) is the integral of a Lagrangian $L$.
Then employing the symmetrized expansion discussed in section 2, we may write\footnote{We are using the fact that any product of matrices may be written as a sum of completely symmetrized products,
where the individual terms in a product must be individual matrices or complete commutators of the form
\[
[X_{i_1}^{(n_1)},[X_{i_2}^{(n_2)},[\dots,[X_{i_{m-1}}^{(n_{m-1})},X_{i_m}^{(n_m)}] \dots ]]]
\] where $(n)$ represents the $n$th time derivative.}
\be
\label{lexp}
L = \hat{L} + \sum_{m=1}^n {1 \over m!} \STr ( L_{(i_1 \ldots i_m)} X^{i_1}\ldots X^{i_m}),
\ee
where the various terms in $\hat{L}$ and $L_{(i_1 \ldots i_n)}$ do not contain any free $X$s. Here, the various individual commutators or differentiated $X$s appearing in a given term of $L$ are to be symmetrized with the remaining $X$s.
Now, for $\delta_\epsilon S$ to vanish under a translation $\delta X = \epsilon$, $\delta L$ must be a total derivative
\be
\label{equal}
\delta_\epsilon L = \epsilon^i {d \over dt} U^i \; .
\ee
Generally, $U^i$ may be written
\[
U^i =  \sum {1 \over m!} \STr ( U^i_{(i_1 \ldots i_m)} X^{i_1}\ldots X^{i_m} ) \; ,
\]
so (\ref{equal}) becomes
\bea
\label{master}
\sum_{m=1}^n {1 \over (m-1)!} \STr (  L_{(i_1 \ldots i_m)} \epsilon^{i_1} X^{i_2} \ldots X^{i_m}) &=& \sum_m {1 \over m!} \STr (  \dot{U}^i_{(i_1 \ldots i_m)} X^{i_1}\ldots X^{i_m} )\\
&& +  {1 \over (m-1)!} \STr(U^i_{(i_1 \ldots i_m)} \dot{X}^{i_1} \ldots X^{i_m} ) \; . \nn
\eea
In this equation, consider the terms with the largest number of free $X$s. To be precise, we can substitute $X \to X + \beta$ and compare the terms with the largest power of $\beta$. Doing this, we find it necessary that
\[
L_{(i_1 \cdots i_n)} = \dot{U}^{i_1}_{i_2 \cdots i_n} + {\cal C}_{i_1 \cdots i_n} \; ,
\]
where ${\cal C}$ is a commutator, ${\cal C} \sim [(X^j)^{(l)}, {\cal A}_j^{i_1 \cdots i_n}]$. Since
\[
\STr([(X^j)^{(l)}, {\cal A}_j^{i_1 \cdots i_n}] X^{i_1} \cdots X^{i_n}) = n \; \STr({\cal A}_j^{i_1 \cdots i_n} [X^{i_1},(X^j)^{(l)}] \cdots X^{i_n}) \; ,
\]
the term in (\ref{lexp}) involving ${\cal C}$ can rewritten such that it has only $n-1$ free $X$s\footnote{This assertion would be incorrect if ${\cal A}$ were $[X^i, X^j]$, but this is impossible, since the $i$ and $j$ indices would have to contract with the indices on two other $X$s which are symmetrized.} and therefore can be absorbed into a redefinition of $L_{i_1 \cdots i_{n-1}}$. With this redefinition, we now have
\be
\label{firstone}
L_{(i_1 \cdots i_n)} = \dot{U}^{i_1}_{i_2 \cdots i_n} \; .
\ee
In particular, $U$ must be completely symmetric on all its indices, so we can write $U^i_{j_1 \cdots j_n} = U_{(i j_1 \cdots j_n)}$.

Comparing the terms in (\ref{master}) with $(n-1)$ free $X$s, we find
\[
L_{(i_1 \cdots i_{n-1})} = U_{(j i_1 \cdots i_{n-1})} \dot{X}^{j} + \dot{U}^{i_1}_{i_2 \cdots i_{n-1}} + {\cal C}_{i_1 \cdots i_{n-1}} \; ,
\]
where ${\cal C}_{i_1 \cdots i_{n-1}}$ is a commutator. As before, by rearranging terms in the trace, we may eliminate ${\cal C}$ in favour of a redefinition of $L_{(i_1 \cdots i_{n-2})}$. Thus,
\[
L_{(i_1 \cdots i_{n-1})} = U_{(j i_1 \cdots i_{n-1})} \dot{X}^{j} + \dot{U}^{i_1}_{i_2 \cdots i_{n-1}} \; ,
\]
and it must be that $U^{i_1}_{i_2 \cdots i_{n-1}}$ is symmetric in all of its indices.
Continuing in this way, we find that by rearranging commutators, it is possible to ensure that all $U$s are completely symmetric tensors and
\be
\label{rest}
L_{(i_1 \cdots i_k)} = U_{(j i_1 \cdots i_k)} \dot{X}^{j} + \dot{U}_{(i_1 \cdots i_k)}
\ee
for all $k \ge 1$. Substituting (\ref{firstone}) and (\ref{rest}) into (\ref{lexp}), it follows that
\[
L = {d \over dt} \sum_n {1 \over n!} \STr(U_{(i_1 \cdots i_n)} X^{i_1} \cdots X^{i_n}) - \tr(U_i \dot{X}^i) + \hat{L} \; ,
\]
where we have integrated by parts to get the second term. Since  $\hat{L}$ and $U^i$ do not contain any free $X$s by assumption, we conclude the action $S$ can be written as the integral of a Lagrangian density with all $X$s in commutators, $\partial_\epsilon \hat{L}(X+ \epsilon)=0$ (up to total derivative terms).

Starting from a Lagrangian $L$ that has been written so that it contains no free $X$s, let us now suppose the action is invariant under $\delta X = \beta t$. Then we must have
\be
\label{invar}
\delta_\beta L = {d \over dt} \beta_i \Phi^i \; .
\ee
for some $\Phi$. We can organize the symmetrized expansions of $L$ and $\Phi$ in terms of the number of free $\dot{X}$s (not appearing in commutators), to write
\bea
\label{lexp2}
L &=& \hat{L} + \sum_{m=1}^n {1 \over m!} \STr ( L_{(i_1 \ldots i_m)} \dot{X}^{i_1}\ldots \dot{X}^{i_m}) \; , \cr
\Phi^i &=& \sum_{m} {1 \over m!} \STr ( \Phi^i_{(i_1 \ldots i_m)} \dot{X}^{i_1}\ldots \dot{X}^{i_m}) \; .
\eea
Here, by assumption, all $X$s and $\dot{X}$s in $\hat{L}$ and $L_{(i_1 \ldots i_m)}$ appear in commutators. In  $\Phi^i_{(i_1 \ldots i_m)}$, all $\dot{X}$s must appear in commutators by assumption. Also, unless we have $\Phi^i \propto X^i$, any free $X$ in $\Phi^i_{(i_1 \ldots i_m)}$ would remain undifferentiated in at least some terms on the right side of (\ref{invar}), and this is not allowed since the left side contains no free $X$s. Thus, either $\Phi^i \propto X^i$ or all $X$s and $\dot{X}$s in $\Phi^i_{(i_1 \ldots i_m)}$ appear in commutators.

Inserting the expansions (\ref{lexp2}) in (\ref{invar}), we have
\bea
\label{master2}
\sum_{m=1}^n {1 \over (m-1)!} \STr (  L_{(i_1 \ldots i_m)} \epsilon^{i_1} \dot{X}^{i_2} \ldots \dot{X}^{i_m}) &=& \sum_m {1 \over m!} \STr (  \dot{\Phi}^i_{(i_1 \ldots i_m)} \dot{X}^{i_1}\ldots \dot{X}^{i_m} )\\
&& +  {1 \over (m-1)!} \STr(\Phi^i_{(i_1 \ldots i_m)} \ddot{X}^{i_1} \ldots \dot{X}^{i_m} ) \; . \nn
\eea
This equation is exactly analogous to (\ref{master}) above, and the rest of the proof proceeds in parallel to that above.\footnote{The only change is to the comment in the previous footnote, which should now deal with three special cases, ${\cal C} \propto [X^i,X^j]$, ${\cal C} \propto [\dot{X}^i, \dot{X^j}]$, and ${\cal C} \propto [X^i,\dot{X}^j]$, for which it is apparently not true that rearranging the commutators
\[
\tr({\cal C}_{i_1 \cdots i_n} \dot{X}^{i_1} \cdots \dot{X}^{i_n})
\]
leads to an expression with no free $X$s and less free $\dot{X}$s. But again, none of these cases are realized. The first two are not possible since the commutator is an antisymmetric rotational tensor whose indices must contract with the indices of symmetrized $\dot{X}$s, while the third is not possible since it would necessarily have an odd number of time derivatives and violate time-reversal invariance.} This time, we end up with the statement that
\[
L = {d \over dt} \sum_n {1 \over n!} \STr(\Phi_{(i_1 \cdots i_n)} \dot{X}^{i_1} \cdots \dot{X}^{i_n}) - \tr(\Phi_i \ddot{X}^i) + \hat{L} \; .
\]
The special case that $\Phi^i \propto X^i$ corresponds to
\[
L_0 = \tr(\dot{X}^2) \; .
\]
Otherwise, all $X$s and $\dot{X}$s in $\Phi_i$ and $\hat{L}$ appear in commutators, so after integrating by parts to remove the first term here, we have succeeded in writing the action as the integral of a Lagrangian density for which all $X$s and $\dot{X}$s appear in commutators.

\section{The matrix vector field $V^\mu(y)$}
\label{navectorfield}

We first discuss the single brane vector field $v^{\mu}(y)$. From the covariance property
under space translations $\delta x^i=a^i$,
\[
\delta v^i = -a^j \partial_j v^i \; , \qquad \delta v^0= -a^j \partial_j v^0 \; ,
\]
follows that $x^i$ and $y^i$ always appear in the combination
$x^i-y^i$. We define the expansion
\bea
v^i & = & \sum_n v^i_{i_1\ldots i_n} (x-y)^{i_1} \ldots (x-y)^{i_n}, \nn \\
v^0 & = & \sum_n v^0_{i_1\ldots i_n} (x-y)^{i_1} \ldots (x-y)^{i_n},
\eea
where the modes $v^{\mu}_{i_1\ldots i_n}$ only contain $\dot{x}$ and higher derivatives.
Using (\ref{vdef}) we can calculate $v^i$ once we know $v^0$
\be
v^i = x^i -y^i + \sum_{p=1}^{\infty} \frac{1}{p!} \frac{d^p x^i}{d t^p} \(v^0\)^p.
\label{vi}
\ee
Furthermore from the defining equations (\ref{vdef}) and (\ref{tdef}) we can calculate
iteratively the modes of $v^0$:
\bea
v^0 & = & 0, \nn \\
v^0_{i_1} & = & \frac{\dot{x}^{i_1}}{1-\dot{x}^2}, \nn \\
v^0_{i_1i_2} & = & \frac{1}{1-\dot{x}^2} \(\frac{\dot{x}^{(i_1}\ddot{x}^{i_2)}}{1-\dot{x}^2}+\frac{3}{2}\frac{\dot{x}^{(i_1}\dot{x}^{i_2)}\dot{x}^k\ddot{x}^k}{(1-\dot{x}^2)^2}\), \nn \\
v^0_{i_1\ldots i_n} & = & \mathcal{O}(x^n).
\label{v0}
\eea
Since the abelian distribution $\theta(y)$ of (\ref{abeliandistr}) forces $y^i=x^i$ we only need
to know the first $n$ modes of $v^{\mu}$ when covariantizing ${d^n}x^i/{dt^n}$ in the manner explained in section \ref{leadingproof2}.

The matrix generalization $V^\mu(y)$ has the form
\be
V^\mu(y)= V^\mu_{\rm sym}(y) + \Delta V^\mu(y),
\label{vNA}
\ee
where $V^\mu_{\rm sym}(y)$ is the expression obtained by replacing all occurrences of $x^i$ in
the expansion of $v^\mu(y)$ with $X^i$ and using the completely symmetrized product of matrices.
The calculation of $\Delta V^\mu(y)$ is much harder than that of the abelian part so we are
forced to work order by order in powers of $X$.
The correction terms to the modes of $\Delta V^\mu(y)$ to fourth order in $X$ are as follows:
\bea
\Delta V^0 & = & \frac{1}{12}\sym \left( \dot{X}^j[\dot{X}^i,[X^j,\dot{X}^i]] + \dot{X}^j[\dot{X}^j,[\dot{X}^i,X^i]] + \dot{X}^j[\dot{X}^i,[\dot{X}^j,X^i]] \right.\nn\\
 & & \left.+\dot{X}^j[X^j,[\ddot{X}^i,X^i]] + \dot{X}^j[\ddot{X}^i,[X^j,X^i]]\right) + \mathcal{O}(X^6), \nn \\
\Delta V^0_{i_1i_2} & = & \frac{1}{12}\sym \left( \dot{X}^j[\dot{X}^{i_1},[\ddot{X}^j,\dot{X}^{i_2}]]
 +\dot{X}^j[\dot{X}^j,[\ddot{X}^{i_1},\dot{X}^{i_2}]] + 2\dot{X}^j[\ddot{X}^{i_1},[\dot{X}^j,\dot{X}^{i_2}]] \right. \nn\\
 & & \left.+ \dot{X}^j[X^j,[\dddot{X}^{i_1},\dot{X}^{i_2}]]
 +\dot{X}^j[\dddot{X}^{i_1},[X^j,\dot{X}^{i_2}]] + \dot{X}^j[\dot{X}^{i_1},[\dot{X}^j,\ddot{X}^{i_2}]]\right.\nn \\
 & & \left. + \dot{X}^j[\ddot{X}^{i_1},[X^j,\ddot{X}^{i_2}]] \right)+ \mathcal{O}(X^6), \nn\\
\Delta V^i_{i_1} & = & \frac{1}{12}\sym \left( \dot{X}^j[\dot{X}^i,[\dot{X}^j,\dot{X}^{i_1}]] + \dot{X}^j[X^j,[\ddot{X}^i,\dot{X}^{i_1}]] + \dot{X}^j[\ddot{X}^i,[X^j,\dot{X}^{i_1}]] \right.\nonumber\\
 & & \left.+\dot{X}^j[\dot{X}^{i_1},[\dot{X}^j,\dot{X}^i]]+\dot{X}^j[X^j,[\ddot{X}^{i_1},\dot{X}^i]] + \dot{X}^j[\ddot{X}^{i_1},[X^j,\dot{X}^i]] \right) + \mathcal{O}(X^6).\nn \\
\eea
Corrections to $\Delta V^i$, $\Delta V^i_{i_1i_2}$, $\Delta V^0_{i_1}$ and $\Delta V^0_{i_1i_2i_3}$
start at order ${\cal O}(X^5)$ and have also been calculated.

Here are some useful expressions for power counting.
\bea
V^0(y)|_{y^i=0}&=&{\cal O}(X^2),\nonumber\\
\d_{i_1}\dots \d_{i_n} V^0(y)|_{y^i=0}&=&{\cal O}(X^n),\nonumber\\
\label{order}
V^i(y)|_{y^i=0}&=&{\cal O}(X),\nonumber\\
\d_jV^i(y)|_{y^i=0}&=&-\delta^{ij} + {\cal O}(X^2),\nonumber\\
\d_{i_1}\dots \d_{i_n} V^i(y)|_{y^i=0}&=&{\cal O}(X^{n+1}).
\eea
For all integers $m \ge 0$,
\bea
\label{count}
\d^{2m} V^0|_{y^i=0} &=& {\cal O}(X^2), \nonumber\\
\d_0 \d^{2m} V^0|_{y^i=0} &=& {\cal O}(X^2), \nonumber\\
\d_j \d^{2m} V^0|_{y^i=0} &=& d^{2m+1} X^j/ dt^{2m+1} + {\cal O}(X^3), \nonumber\\
\d^{2m} V^i|_{y^i=0} &=& -d^{2m} X^i/ dt^{2m} + {\cal O}(X^3), \nonumber\\
\d_0 \d^{2m} V^i|_{y^i=0} &=& -d^{2m+1}X^i/ dt^{2m+1} + {\cal O}(X^3), \nonumber\\
\d_j\d^{2m} V^i|_{y^i=0} &=& \delta^{ij} \delta_{m,0} + {\cal O}(X^2).
\eea
Here $\d^2 \equiv \eta^{\mu\nu} \d_\mu \d_\nu$, $\eta = \diag(-1, 1, \dots , 1)$.

\section{Calculation of the 2-commutator corrections to the D0-brane currents and covariant matrix distribution}

In this appendix, we outline the calculation of the two-commutator corrections to the abelian expressions for the D0-brane current and covariant matrix distribution,
required by demanding Lorentz covariance.
It is convenient to begin with the D0-brane current, since parts of this calculation will appear again when we discuss the covariant matrix distribution.

\subsection{Outline of the calculation for the D0-brane current}
\label{2commcurrentcalc}

To calculate corrections to the currents, we use $\rho=\rho_{(0)}+\rho_{(1)}+{\cal O}([\cdot,\cdot]^4)$ and $J^i=J^i_{(0)}+J^i_{(1)}+{\cal O}([\cdot,\cdot]^4)$ in eq.~(\ref{cons}).
The zeroth orders $\rho_{(0)}$ and $J^i_{(0)}$ are given by eq.~(\ref{leadcurr2}) and we use (\ref{nonabelrules},\ref{2commtranslawterm})
as the transformation law. We can work this out up to two commutators using (\ref{nestedsym}). The thus calculated constraints can be further simplified if we make the ansatz
\bea
\rho^{(i_1 \cdots i_n)} & = & \STr \left[ \sum_p C_n^{(i_1\ldots i_p} X^{i_{p+1}} \ldots X^{i_n)} \right], \\
J^{i;(i_1 \cdots i_n)} & = & \STr \left[ \sum_p \left[D_n^{(i_1\ldots i_p} \dot{X}^{|i|} +E_n^{i;(i_1\ldots i_p} \right] X^{i_{p+1}} \ldots X^{i_n)} \right],
\eea
where $C_n=D_n=E_n=0$ if $n<p$.
From covariance of the current under spatial translations, the first equation of (\ref{nonabelrules}), we find that
\be
(n-p) C_n^{(p)} = n C_{n-1}^{(p)}, \qquad (n-p) D_n^{(p)} = n D_{n-1}^{(p)}, \qquad (n-p) E_n^{(p)} = n E_{n-1}^{(p)},
\ee
This allows us to write everything in terms of $C_p,D_p,E_p$ thereby producing the combinatorial factors in (\ref{currenttemplate}).
In fact, from the terms containing $\b^i$ we find $D_n=C_n$ and current conservation leads to (\ref{currentcons2}).

In the following we try to find a solution which is covariant without using the cyclicity of the trace.
In terms of $C_p,E_p$ the constraints read
\bea
& & p \, \b^l \delta^p_1 T^{i_1l} + \frac{p(p-1)}{12} \b^l \delta^p_2 N^{(i_1i_2)l}
- \frac{p(p-1)(p-2)}{12} \b^l \delta^p_3 M^{(i_1i_2i_3)l} \nn \\
& & + \delta'_\b C_p^{(i_1 \ldots i_p)}
- \b^l E_p^{l;(i_1 \ldots i_p)}
+ \b_l \frac{d}{dt} C_{p+1}^{(li_1 \ldots i_{p})} = 0, \nn \\
& & \b^l \delta^p_0 \frac{d}{dt} T^{il} + \frac{p}{12} \b^l \delta^p_1 K^{i;i_1l}
- \frac{p(p-1)}{12} \b^l \delta^p_2 L^{i;(i_1i_2)l} \nn \\
& & + \delta_\b'{}^{(+1)} E_p^{i;(i_1 \ldots i_p)}
+ \b^l C_{p+1}^{(li_1 \ldots i_{p})} \ddot{X}^{i} + \b^l \frac{d}{dt} E_{p+1}^{i;(l i_1 \ldots i_{p})} = 0,
\label{simcurrentconstr}
\eea
with $T^{il}$ given by (\ref{2commtranslawterm}), $\delta'$ by (\ref{deltap}) and we furthermore defined
\be
\delta_\b'{}^{(m)} A^{r_1 \ldots r_s} = \delta'_\b A^{r_1 \ldots r_s} + m \, \b^l \dot{X}^l A^{r_1 \ldots r_s},
\label{deltap2}
\ee
and
\bea
N^{(i_1i_2)j} & = & [\dot{X}^{(i_1},[X^{i_2)},X^j]] + [X^j,[X^{(i_1},\dot{X}^{i_2)}]],\nn \\
M^{(i_1i_2i_3)j} & = & [X^{(i_1},\dot{X}^{i_2}][X^{i_3)},X^j], \nn \\
K^{i;i_1j} & = & [\ddot{X}^{i},[X^{i_1},X^j]] + [\dot{X}^{i_1},[\dot{X}^{i},X^j]] + [\dot{X}^{i},[X^{i_1},\dot{X}^j]] \nn \\
& & + [X^{j},[X^{i_1},\ddot{X}^i]] + [X^j,[\dot{X}^i,\dot{X}^{i_1}]] + [\dot{X}^{j},[X^{i_1},\dot{X}^i]], \nn \\
L^{i;(i_1i_2)j} & = & [X^{(i_1},\ddot{X}^{|i|}][X^{i_2)},X^j] + [X^{(i_1},\dot{X}^{|i|}][X^{i_2)},\dot{X}^j] +
[\dot{X}^{i},\dot{X}^{(i_1}][X^{i_2)},X^j] \nn \\
& & + [X^{(i_1},\dot{X}^{i_2)}][\dot{X}^{i},X^j].
\label{constensors1}
\eea

Interestingly if we apply a field redefinition (\ref{FR}) the boost transformation law
will change as
\be
\Delta T^{ij} = [\delta_\b,\tilde{\delta}]X^i = \delta' F^i,
\ee
where we defined $\tilde{\delta}X^i=F^i$.
As is clear from the first eq.\ in (\ref{simcurrentconstr}) for $p=1$, this can be completely absorbed
by putting $C^i_1 \rightarrow C^i_1 - F^i$. In particular, we can apply a field redefinition to put $C^i_1=0$ but
that will make the transformation law (\ref{2commtranslawterm}) more complicated.

Solving these equations is tedious and requires heavy use of formula (\ref{varterm}).
We present the result in subsection \ref{2commresults}.

\subsection{Outline of the calculation for the covariant matrix distribution}
\label{2commdistrcalc}

To make optimal use of the ansatz (\ref{sqrtansatz}) we can apply the following formula
\bea
\lefteqn{\delta'\left[\frac{\d}{\d \dot{X}_{j_1}} \ldots\frac{\d}{\d \dot{X}_{j_n}}\sqrt{1-\dot{X^i}\dot{X^i}}\, A^{(i_1\ldots i_p);(j_1 \ldots j_n)}\right]=} \nn \\
& = & \frac{\d}{\d \dot{X}_{j_1}} \ldots\frac{\d}{\d \dot{X}_{j_n}}\sqrt{1-\dot{X^i}\dot{X^i}} \, \delta'^{(n-1)} A^{(i_1\ldots i_p);(j_1 \ldots j_n)}\nn \\
& & + n(n-2) \b_l \frac{\d}{\d \dot{X}_{j_2}} \ldots\frac{\d}{\d \dot{X}_{j_n}}\sqrt{1-\dot{X^i}\dot{X^i}} \, A^{(i_1 \ldots i_p);(l j_2 \ldots j_n)},
\eea
with $\delta'$ given by (\ref{deltap}) and $\delta'^{(m)}$ by (\ref{deltap2}).
The constraint (\ref{constraintdistr}) then reduces for $C_n$ and $E_n$ to (\ref{simcurrentconstr}) and for the $R$ to
\bea
& & \b^l \frac{1}{12} \delta^p_0 \delta^k_2 U^{(j_1j_2)l} - \b^l \frac{1}{12} \delta^p_0 \delta^k_3 Q^{(j_1j_2j_3)l}
-\b^l \frac{p}{12} \delta^p_1 \delta^k_2 V^{(j_1j_2);i_1l} \nn \\
& & + \delta' R_p^{(j_1\ldots j_k);(i_1 \ldots i_p)}
+(k+1)(k-1) R_p^{(lj_1\ldots j_k);(i_1 \ldots i_p)} + \ddot{X}^{(j_1} R_{p+1}^{j_2\ldots j_k);(li_1 \ldots i_p)} \nn \\
& & + \frac{d}{dt}R_{p+1}^{(j_1\ldots j_k);(li_1 \ldots i_p)} =0,
\eea
with
\bea
U^{(ik)j} & = & [\ddot{X}^{(i},[\dot{X}^{k)},X^j]] + [X^j,[\dot{X}^{(i},\ddot{X}^{k)}]]+[\dot{X}^{(i},[\dot{X}^{k)},\dot{X}^j]] \nn \\
V^{(ik);i_1j} & = & [\dot{X}^{(i},\ddot{X}^{k)}][X^{i_1},X^j]+[\dot{X}^{(i},X^{|j}][X^{i_1|},\ddot{X}^{k)}]
+ [\dot{X}^{(i},\dot{X}^{|j}][X^{i_1|},\dot{X}^{k)}] \nn \\
& & + [\dot{X}^{(i},\dot{X}^{|i_1|}][\dot{X}^{k)},X^{j}] \nn \\
Q^{(ikm)j} & = & [\dot{X}^{(i},\ddot{X}^k][\dot{X}^{m)},X^j].
\eea
Our results for all of these tensors up to two commutators are given below.

\subsection{Results}
\label{2commresults}

Note that since we do not use the cyclicity of the trace the solutions will split into a part containing nested commutators (as in the
tensors $N$ and $K$ in (\ref{constensors1})) and a part containing unnested commutators (as in the tensors $M$ and $L$ in (\ref{constensors1})).
A minimal but non-unique solution of the constraints (\ref{simcurrentconstr}) using the transformation law (\ref{2commtranslawterm}) is given by:
\bea
\lefteqn{C_{3,\text{unnested}}^{(i_1i_2i_3)}=} \nn \\
& &\frac{1}{2} E^{jt} \dot{X}_t M^{(i_1 i_2 i_3)}{}_j
+ \frac{1}{4} E^{j_1t_1} \dot{X}_{t_1} E^{j_2t_2} \dot{X}_{t_2} \ddot{X}^{(i_1}[X^{i_2},X_{j_1}][X^{i_3)},X_{j_2}], \nn \\
\lefteqn{C_{2,\text{nested}}^{(i_1i_2)}=} \nn \\
& & -\frac{1}{6} E^{jt} \dot{X}_t N^{(i_1 i_2)}{}_j
-\frac{1}{6} E^{j_1t_1} \dot{X}_{t_1} E^{j_2t_2} \dot{X}_{t_2} \ddot{X}^{(i_1}[X_{j_1},[X^{i_2)},X_{j_2}]], \nn \\
\lefteqn{C_{2,\text{unnested}}^{(i_1i_2)} =} \nn \\
& & -\frac{1}{12}E^{j_1t_1} \dot{X}_{t_1} E^{j_2t_2} \dot{X}_{t_2}
\Big( 2 [X_{j_1}, \ddot{X}^{(i_1}][X^{i_2)},X_{j_2}] + 2 [X^{(i_1}, \dot{X}^{i_2)}][X_{j_1},\dot{X}_{j_2}] \nn \\
& & +[X_{j_1},\dot{X}^{(i_1}][X_{j_2},\dot{X}^{i_2)}]+2[\dot{X}^{(i_1},\dot{X}_{j_1}][X^{i_2)},X_{j_2}]\Big) \nn \\
& & -\frac{1}{3 }E^{j_1t_1} \dot{X}_{t_1} E^{j_2t_2} \dot{X}_{t_2} E^{j_3t_3}\dot{X}_{t_3} \ddot{X}^{(i_1} [X_{j_1},\dot{X}_{j_2}][X^{i_2)},X_{j_3}] \nn \\
& & - \frac{1}{6} E^{jt} \ddot{X}_t \left(1+\dot{X}_u E^{uv}\dot{X}_v\right)E^{j_2t_2}\dot{X}_{t_2} \Big(
(h_1+h_3) [X^{(i_1},\dot{X}_{j_2}][X^{i_2)},X_j] \nn \\
& & +(h_1+h_2) [X^{(i_1},\dot{X}^{i_2)}][X_{j_2},X_j]
+(h_2+h_3) [X_{j_2},\dot{X}^{(i_1}][X^{i_2)},X_j] \nn \\
& & +h_4 [X_{j},\dot{X}^{(i_2}][X_{j_2},X^{i_1)}]
+ h_5 [X^{(i_2},\dot{X}_{j}][X_{j_2},X^{i_1)}]\Big)\nn \\
& & - \frac{1}{6} E^{jt} \ddot{X}_t \left(1+\dot{X}_u E^{uv}\dot{X}_v\right) E^{j_2t_2} \dot{X}_{t_2} E^{j_3t_3} \dot{X}_{t_3}\Big(
(h_1+h_3) \ddot{X}_{j_2}[X^{(i_1},X_{j_3}][X^{i_2)},X_j] \nn \\
& & + (h_1+h_2+h_4) \ddot{X}^{(i_1}[X^{i_2)},X_{j_2}][X_{j_3},X_j]
+h_5 \ddot{X}_j[X^{(i_1},X_{j_2}][X_{j_3},X^{i_1)}]\Big), \nn \\
\lefteqn{C^{i_1}_{1,\text{nested}} =} \nn \\
& & \frac{1}{24} E^{j_1 t_1} \dot{X}_{t_1} E^{j_2 t_2} \dot{X}_{t_2}
\Big( - [X_{j_1},[\ddot{X}^{i_1},X_{j_2}]]
-2 [\dot{X}^{i_1},[\dot{X}_{j_1},X_{j_2}]]
+ [\dot{X}_{j_1},[\dot{X}_{i_1},X_{j_2}]]\Big) \nn \\
& & + \frac{1}{24} E^{j_1 t_1} \dot{X}_{t_1} E^{j_2 t_2} \dot{X}_{t_2} E^{j_3 t_3} \dot{X}_{t_3}
\Big( 2 \ddot{X}^{i_1}[X_{j_1},[X_{j_2},\dot{X}_{j_3}]] - \ddot{X}_{j_1} [X_{j_2},[X_{j_3},\dot{X}^{i_1}]]
\Big) \nn \\
& & + \frac{1}{12} E^{jt} \ddot{X}_t \(1+\dot{X}_u E^{uv}\dot{X}_v\) E^{j_2t_2}\dot{X}_{t_2} \Big(
f_1 [\dot{X}^{i_1},[X^{j_2},X_j]] + f_2 [\dot{X}_{j_2},[X^{i_1},X_j]] \nn \\
& & + f_3 [X_j,[X^{i_1},\dot{X}_{j_2}]] + f_4 [X_j,[X_{j_2},\dot{X}^{i_1}]] + f_5 [\dot{X}_j,[X^{i_1},X_{j_2}]]\Big) \nn \\
& & \frac{1}{12} E^{jt} \ddot{X}_t \left(1+\dot{X}_u E^{uv}\dot{X}_v\right) E^{j_2t_2} \dot{X}_{t_2} E^{j_3t_3} \dot{X}_{t_3} \Big(
f_1 \ddot{X}^{i_1} [X_{j_2},[X_{j_3},X_j]] \nn \\
& & + f_2 \ddot{X}_{j_2} [X_{j_3},[X^{i_1},X_j]]
+ f_3 \ddot{X}_{j_2} [X_j,[X^{i_1},X_{j_3}]] + f_5 \ddot{X}_{j} [X_{j_2},[X^{i_1},X_{j_3}]] \Big), \nn \\
\lefteqn{C_{1,\text{unnested}}^{i_1} = } \nn \\
& & \frac{1}{8} E^{j_1t_1} \dot{X}_{t_1} E^{j_2t_2} \dot{X}_{t_2} E^{j_3t_3} \dot{X}_{t_3}
\([\ddot{X}^{i_1},X_{j_1}][\dot{X}_{j_2},X_{j_3}] + [\dot{X}^{i_1},\dot{X}_{j_1}][X_{j_2},\dot{X}_{j_3}]\) \nn \\
& & + \frac{3}{16} E^{j_1t_1} \dot{X}_{t_1} E^{j_2t_2} \dot{X}_{t_2} E^{j_3t_3} \dot{X}_{t_3} E^{j_4t_4} \dot{X}_{t_4}
\ddot{X}^{i_1} [\dot{X}_{j_1},X_{j_2}][\dot{X}_{j_3},X_{j_4}] \nn \\
& & + \frac{1}{12} E^{jt} \ddot{X}_t \left(1+\dot{X}_u E^{uv}\dot{X}_v\right)E^{j_2t_2}\dot{X}_{t_2} E^{j_3t_3}\dot{X}_{t_3} \nn \\
& & \Bigg(\(h_1-h_2+f_1-f_2-f_3+f_4\) [\dot{X}^{i_1},\dot{X}_{j_2}][X_{j_3},X_j]
+\(h_1-f_3\) [X^{i_1},\ddot{X}_{j_2}][X_{j_3},X_j] \nn \\
& & +\(h_1+h_4-2f_3\) [X^{i_1},\dot{X}_{j_2}][\dot{X}_{j_3},X_j]
+ \(h_1-h_5-f_5\) [X^{i_1},\dot{X}_{j_2}][X_{j_3},\dot{X}_j] \nn \\
& & +\(-h_2+f_4-1\) [\ddot{X}^{i_1},\dot{X}_{j_2}][X_{j_3},X_j]
+\(-h_2+h_4+f_1+2f_4\) [\dot{X}^{i_1},X_{j_2}][\dot{X}_{j_3},X_j] \nn \\
& & +\(-h_2-h_5+f_1-f_5-1\) [\dot{X}^{i_1},X_{j_2}][X_{j_3},\dot{X}_j]
+h_3 [X^{i_1},X_{j}][X_{j_2},\ddot{X}_{j_3}] \nn \\
& & +\(h_3-f_2-f_3-f_4+1\) [\dot{X}^{i_1},X_{j}][X_{j_2},\dot{X}_{j_3}]
+\(h_3-f_2\) [X^{i_1},\dot{X}_{j}][X_{j_2},\dot{X}_{j_3}] \nn \\
& & +\(-h_4+h_5-f_2+f_5\) [X^{i_1},{X}_{j_2}][\dot{X}_{j},\dot{X}_{j_3}]
+\(-h_4+f_3\) [X^{i_1},X_{j_2}][X_{j},\ddot{X}_{j_3}] \nn \\
& & -h_5 [{X}^{i_1},{X}_{j_2}][X_{j_3},\ddot{X}_j] \Bigg) \nn \\
& & + \frac{1}{12} E^{jt} \ddot{X}_t \left(1+\dot{X}_u E^{uv}\dot{X}_v\right)E^{j_2t_2}\dot{X}_{t_2} E^{j_3t_3}\dot{X}_{t_3} E^{j_4t_4}\dot{X}_{t_4}\nn \\
& & \Bigg( \(-f_3+h_1+h_4\) \dddot{X}_{j_2} [X^{i_1},X_{j_3}][X_{j_4},X_j] \nn \\
& & +\(f_1-f_2-f_3-f_4 + 3\) \ddot{X}^{i_1} [X_{j_2},\dot{X}_{j_3}][X_{j_4},X_j] \nn \\
& & +\(-f_2+f_5+h_3-h_4-h_5\) \ddot{X}_{j} [X_{j_2},\dot{X}_{j_3}][X^{i_1},X_{j_4}]  \nn \\
& & + \(-f_1+f_2+f_3-2f_4 -h_1+5h_2-h_4\) \ddot{X}_{j_2} [X_{j_3},\dot{X}^{i_1}][X_{j_4},X_j] \nn \\
& & + \(f_2-f_5 +h_1+h_4-5h_5\) \ddot{X}_{j_2} [X_{j_3},\dot{X}_{j}][X^{i_1},X_{j_4}] \nn \\
& & + \(3f_3-h_1-6h_4\) \ddot{X}_{j_2} [X_{j},\dot{X}_{j_3}][X^{i_1},X_{j_4}] \nn \\
& & + \(3f_3-6h_1-h_4\) \ddot{X}_{j_2} [X_{j},X_{j_3}][X^{i_1},\dot{X}_{j_4}]
+ 5 h_3 \ddot{X}_{j_2} [X_{j_3},\dot{X}_{j_4}][X^{i_1},X_j] \Bigg) \nn \\
& & + \frac{1}{12} E^{jt} \ddot{X}_t \left(1+\dot{X}_u E^{uv}\dot{X}_v\right)E^{j_2t_2}\dot{X}_{t_2} E^{j_3t_3}\dot{X}_{t_3} E^{j_4t_4}\dot{X}_{t_4} E^{j_5t_5}\dot{X}_{t_5} \nn \\
& & \(3f_3 - 6h_1 -6 h_4\) \ddot{X}_{j_2} \ddot{X}_{j_3} [X_j,X_{j_4}][X^{i_1},X_{j_5}] \nn \\
& & + \frac{1}{12} E^{jt} \dddot{X}_t \left(1+\dot{X}_u E^{uv}\dot{X}_v\right)E^{j_2t_2}\dot{X}_{t_2} E^{j_3t_3}\dot{X}_{t_3} \nn \\
& & \Bigg( h_1 [X^{i_1},\dot{X}_{j_2}][X_{j_3},X_j] + h_2 [X_{j_2},\dot{X}^{i_1}][X_{j_3},X_j] + h_3 [X_{j_2},\dot{X}_{j_3}][X^{i_1},X_j] \nn \\
& & + h_4 [X^{i_1},X_{j_2}][\dot{X}_{j_3},X_j] + h_5 [X_{j_2},X^{i_1}][X_{j_3},\dot{X}_j] \Bigg) \nn \\
& & + \frac{1}{12} E^{jt} \dddot{X}_t \left(1+\dot{X}_u E^{uv}\dot{X}_v\right)E^{j_2t_2}\dot{X}_{t_2} E^{j_3t_3}\dot{X}_{t_3} E^{j_4t_4}\dot{X}_{t_4}\nn \\
& & \(h_1+h_4\) \ddot{X}_{j_2} [X^{i_1},X_{j_3}][X_{j_4},X_j] \nn \\
& & + \frac{1}{12} E^{j_1t_1} \ddot{X}_{t_1} E^{j_2t_2}\ddot{X}_{t_2} \left(1+\dot{X}_u E^{uv}\dot{X}_v\right)^2 E^{j_3t_3}\dot{X}_{t_3} E^{j_4t_4}\dot{X}_{t_4}\nn \\
& & \Bigg( \(h_1+h_3\) \ddot{X}_{j_3} [X^{i_1},X_{j_1}][X_{j_4},X_{j_2}] + h_2 \ddot{X}^{i_1} [X_{j_3},X_{j_1}][X_{j_4},X_{j_2}] \nn \\
& & + h_5 \ddot{X}_{j_1} [X_{j_3},X_{j_2}][X_{j_4},X^{i_1}] \Bigg), \nn \\
\lefteqn{E_{2,\text{unnested}}^{i;(i_1i_2)} =} \nn \\
& & \frac{1}{6} E^{jt} \dot{X}_t L^{i;(i_1i_2)}{}_j
+\frac{1}{6} E^{j_1t_1} \dot{X}_{t_1} E^{j_2t_2} \dot{X}_{t_2} \Big( \frac{1}{2} \dddot{X}\!{}^i[X^{(i_1},X_{j_1}][X^{i_2)},X_{j_2}] \nn \\
& & + \ddot{X}^i[X^{(i_1},\dot{X}_{j_1}][X^{i_2)},X_{j_2}]
+ \ddot{X}_{j_1}[X^{(i_1},\dot{X}^{|i|}][X^{i_2)},X_{j_2}]
+ \ddot{X}^{(i_1}[\dot{X}^{|i|},X_{j_1}][X^{i_2)},X_{j_2}]\Big) \nn \\
& & + \frac{1}{6} E^{j_1t_1} \dot{X}_{t_1} E^{j_2t_2} \dot{X}_{t_2} E^{j_3t_3} \dot{X}_{t_3}
\ddot{X}^i\ddot{X}_{j_1} [X^{(i_1},X_{j_2}][X^{i_2)},X_{j_3}] \nn \\
& & + \frac{1}{6} E^{jt} \ddot{X}_t \left(1+\dot{X}_u E^{uv}\dot{X}_v\right) \Big(
h_1 [X^{i},\dot{X}^{(i_1}][X^{i_2)},X_j]+h_2 [X^{(i_1},\dot{X}^{|i|}][X^{i_2)},X_j] \nn \\
& & +h_3 [X^{(i_1},\dot{X}^{i_2)}][X^{i},X_j]
 +h_4 [X_{j},\dot{X}^{(i_2}][X^{i_1)},X^i] + h_5 [X^{(i_1},\dot{X}_{j}][X^{i_2)},X^i]\Big)\nn \\
& & + \frac{1}{6} E^{jt} \ddot{X}_t \left(1+\dot{X}_u E^{uv}\dot{X}_v\right) E^{j_2t_2} \dot{X}_{t_2} \Big(
h_1 \ddot{X}^{(i_1}[X^{|i|},X_{j_2}][X^{i_2)},X_j] \nn \\
& & + h_2 \ddot{X}^{i}[X^{(i_1},X_{j_2}][X^{i_2)},X_j]
h_3 \ddot{X}^{(i_1}[X^{i_2)},X_{j_2}][X^{i},X_j]
+h_4 \ddot{X}^{(i_1}[X_j,X_{j_2}][X^{i_2)},X^i] \nn \\
& & +h_5 \ddot{X}_j[X^{(i_1},X_{j_2}][X^{i_2)},X^i]\Big),
\label{currentsol}
\eea
where $f_1,f_2,f_3,f_4,f_5,h_1,h_2,h_3,h_4,h_5$ are arbitrary coefficients.
We have current conservation (\ref{currentcons2}) if these satisfy
\be
f_1+f_2=1, \qquad f_3+f_4=1, \qquad h_1+h_2+h_3=1.
\ee
Furthermore $C_0=1$ --- this produces the zeroth order result (\ref{leadcurr2}) --- and again from
current conservation $E^i_0=\frac{d}{dt}C^i_1$. If all the other tensors are known we can straightforwardly find
$E_1^{i;i_1}$ from the first equation in (\ref{simcurrentconstr}) for $p=1$. All other tensors $C_p$ for $p \ge 4$
and $E_p$ for $p \ge 3$ are zero.

A minimal but non-unique solution for the $R$ tensors appearing in the covariant matrix distribution is given by
\bea
\lefteqn{R_0^{(ik)} = } \nn \\
& & - \frac{1}{12} E^{jt} \dot{X}_t U^{ik}{}_j -\frac{1}{12} E^{j_1t_1} \dot{X}_{t_1} E^{j_2t_2} \dot{X}_{t_2}\Big( \dddot{X}\!{}^{(i}[X_{j_1},[\dot{X}^{k)},X_{j_2}]] \nn \\
& & + \ddot{X}^{(i}[\dot{X}_{j_1},[\dot{X}^{k)},X_{j_2}]] + \ddot{X}_{j_1}[\dot{X}^{(i},[\dot{X}^{k)},X_{j_2}]]+ \ddot{X}^{(i}[X_{j_1},[\dot{X}^{k)},\dot{X}_{j_2}]] \Big) \nn \\
& & -\frac{1}{6} E^{j_1t_1} \dot{X}_{t_1} E^{j_2t_2} \dot{X}_{t_2} E^{j_3t_3} \dot{X}_{t_3} \ddot{X^{(i}} \ddot{X}_{j_1} [X_{j_2},[\dot{X}^{k)},X_{j_3}]] \nn\\
& & - \frac{1}{12} E^{j_1t_1} \dot{X}_{t_1} E^{j_2t_2} \dot{X}_{t_2}
\Big( [\dot{X}^{(i},X_{j_1}][X_{j_2},\dddot{X}\!{}^{k)}] + 2[\dot{X}^{(i},\ddot{X}^{k)}][X_{j_1},\dot{X}_{j_2}] \nn \\
& & +\frac{1}{2} [\ddot{X}^{(i},X_{j_1}][\ddot{X}^{k)},X_{j_2}] + [\ddot{X}^{(i},X_{j_1}][\dot{X}_{j_2},\dot{X}^{k)}]
+ [\ddot{X}^{(i},\dot{X}_{j_1}][\dot{X}^{k)},X_{j_2}] \nn \\
& & + [\dot{X}^{(i},\ddot{X}_{j_1}][\dot{X}^{k)},X_{j_2}] + \frac{3}{2} [\dot{X}^{(i},\dot{X}_{j_1}][\dot{X}^{k)},\dot{X}_{j_2}] \Big) \nn \\
& & + \frac{1}{12} E^{jt} \ddot{X}_t \left(1+\dot{X}_u E^{uv}\dot{X}_v\right) E^{j_2t_2} \dot{X}_{t_2} \ddot{X}^{(k} \nn \\
& & \Big(f_1 [\dot{X}^{i)},[X_{j_2},X_j]] + f_2 [\dot{X}_{j_2},[X^{i)},X_j]] \nn \\
& & + f_3 [X_j,[X^{i)},\dot{X}_{j_2}]] + f_4 [X_j,[X_{j_2},\dot{X}^{i)}]] + f_5 [\dot{X}_j,[X^{i)},X_{j_2}]]\Big) \nn \\
& & + \frac{1}{12} E^{jt} \ddot{X}_t \left(1+\dot{X}_u E^{uv}\dot{X}_v\right) E^{j_2t_2} \dot{X}_{t_2} E^{j_3t_3} \dot{X}_{t_3} \ddot{X}^{(k} \nn \\
& & \Big(f_1 \ddot{X}^{i)} [X_{j_2},[X_{j_3},X_j]]
+ f_2 \ddot{X}_{j_2} [X_{j_3},[X^{i)},X_j]] \nn \\
& & + f_3 \ddot{X}_{j_2} [X_j,[X^{i)},X_{j_3}]]
+ f_5 \ddot{X}_{j} [X_{j_2},[X^{i)},X_{j_3}]] \Big), \nn \\
& & - \frac{1}{12} E^{jt} \ddot{X}_t \left(1+\dot{X}_u E^{uv}\dot{X}_v\right) E^{j_2t_2} \dot{X}_{t_2} V^{ik;}{}_{j_2j} \nn \\
& & - \frac{1}{12} E^{jt} \ddot{X}_t \left(1+\dot{X}_u E^{uv}\dot{X}_v\right) E^{j_2t_2} \dot{X}_{t_2} E^{j_3t_3} \dot{X}_{t_3} \nn \\
& & \Bigg( \dddot{X^{(i}} \Big((1+h_2)[\dot{X}^{k)},X_{j_2}][X_{j_3},X_j]-h_1[X^{k)},\dot{X}_{j_2}][X_{j_3},X_j]
-h_3 [X^{k)},X_{j}][X_{j_2},\dot{X}_{j_3}] \nn \\
& & -h_4 [X^{k)},X_{j_2}][\dot{X}_{j_3},X_j] - h_5[X^{k)},X_{j_2}][\dot{X}_{j},X_{j_3}] \Big)\nn \\
& & + \ddot{X}^{(i}\Big( (f_4-2h_2)[X_{j_2},\ddot{X}^{k)}][X_{j_3},X_j] \nn \\
& & +(1-f_1+f_2+f_3-f_4-2h_1+2h_2) [\dot{X}^{k)},\dot{X}_{j_2}][X_{j_3},X_j] \nn \\
& & +(-1+f_1-f_5-2h_2-2h_5) [\dot{X}^{k)},X_{j_2}][\dot{X}_{j},X_{j_3}] \nn \\
& & +(1+f_2+f_3+f_4-2h_3) [\dot{X}^{k)},X_{j}][X_{j_2},\dot{X}_{j_3}] \nn \\
& & + (f_3-2h_1) [X^{k)},\ddot{X}_{j_2}][X_{j_3},X_j]
+2(f_3-h_2-h_4) [X^{k)},\dot{X}_{j_2}][\dot{X}_{j_3},X_j] \nn \\
& & +(f_5-2h_1+2h_5) [X^{k)},\dot{X}_{j_2}][X_{j_3},\dot{X}_j]
+ (f_1 +2 f_4-2h_2+2h_4) [\dot{X}^{k)},X_{j_2}][X_{j},\dot{X}_{j_3}] \nn \\
& & +(f_2-2h_3) [X^{k)},\dot{X}_{j}][X_{j_2},\dot{X}_{j_3}]
-2h_3 [X^{k)},X_{j}][X_{j_2},\ddot{X}_{j_3}] \nn \\
& & + (f_3-2h_4) [X^{k)},X_{j_2}][\ddot{X}_{j_3},X_{j}]
+ (-f_2+f_5-2h_4+2h_5) [X^{k)},X_{j_2}][\dot{X}_{j_3},\dot{X}_{j}] \nn \\
& & -2 h_5 [X^{k)},{X}_{j_2}][\ddot{X}_{j},X_{j_3}] \Big)
+ \ddot{X}_j [\dot{X}^{(i},X_{j_2}][X_{j_3},X^{k)}] \Bigg) \nn \\
& & - \frac{1}{12} E^{jt} \ddot{X}_t \left(1+\dot{X}_u E^{uv}\dot{X}_v\right) E^{j_2t_2} \dot{X}_{t_2} E^{j_3t_3} \dot{X}_{t_3} E^{j_4t_4} \dot{X}_{t_4}\nn \\
& & \Big( (f_3-2(h_1+h_4)) \dddot{X}_{j_2} \ddot{X}^{(i}  [X^{k)},X_{j_3}][X_{j_4},X_{j}]
-(h_1+h_4) \dddot{X}\!{}^{(i} \ddot{X}_{j_2}  [X^{k)},X_{j_3}][X_{j_4},X_{j}] \nn \\
& & (f_2 -f_5-2h_3+2h_4+2h_5) \ddot{X}^{(i} \ddot{X}_j [X^{k)},X_{j_2}][X_{j_3},\dot{X}_{j_4}] \nn \\
& & +(2 - f_1 + f_2 + f_3 - 2f_4-2h_1+10h_2-2h_4) \ddot{X}^{(i} \ddot{X}_{j_2} [\dot{X}^{k)},X_{j_3}][X_{j_4},X_{j}] \nn \\
& & +(3 f_3-12h_1-2h_4) \ddot{X}^{(i} \ddot{X}_{j_2} [X^{k)},\dot{X}_{j_3}][X_{j_4},X_{j}]
- 10 h_3 \ddot{X}^{(i} \ddot{X}_{j_2} [X^{k)},X_{j}][X_{j_3},\dot{X}_{j_4}] \nn \\
& & +(3 f_3-2h_1-12h_4) \ddot{X}^{(i} \ddot{X}_{j_2} [X^{k)},X_{j_3}][\dot{X}_{j_4},X_{j}] \nn \\
& & +(-f_2+f_5-2h_1-2h_4+10h_5) \ddot{X}^{(i} \ddot{X}_{j_2} [X^{k)},X_{j_3}][X_{j_4},\dot{X}_{j}] \nn \\
& & +(-1 - f_1 +f_2 +f_3 +f_4) \ddot{X}^{(i} \ddot{X}^{k)} [X_{j_2},\dot{X}_{j_3}][X_{j_4},X_{j}] \Big) \nn \\
& & - \frac{1}{4} E^{jt} \ddot{X}_t \left(1+\dot{X}_u E^{uv}\dot{X}_v\right) E^{j_2t_2} \dot{X}_{t_2} E^{j_3t_3} \dot{X}_{t_3} E^{j_4t_4} \dot{X}_{t_4}E^{j_5t_5} \dot{X}_{t_5}\nn \\
& & (f_3-4(h_1+h_4)) \ddot{X}^{(i} \ddot{X}_{j_2} \ddot{X}_{j_3} [X^{k)},X_{j_4}][X_{j_5},X_{j}] \nn \\
& & +\frac{1}{6} E^{jt} \dddot{X}_t \left(1+\dot{X}_u E^{uv}\dot{X}_v\right) E^{j_2t_2} \dot{X}_{t_2} E^{j_3t_3} \dot{X}_{t_3} \ddot{X}^{(k} \nn \\
& & \Big( h_1 [X^{i)},\dot{X}_{j_2}][X_{j_3},X_j]+h_2 [X_{j_2},\dot{X}^{i)}][X_{j_3},X_j] \nn \\
& & +h_3 [X_{j_2},\dot{X}_{j_3}][X^{i)},X_j]
 +h_4 [X_{j},\dot{X}_{j_2}][X_{j_3},X^{i)}] + h_5 [X_{j_2},\dot{X}_{j}][X_{j_3},X^{i)}]\Big)\nn \\
& & +\frac{1}{6} E^{jt} \dddot{X}_t \left(1+\dot{X}_u E^{uv}\dot{X}_v\right) E^{j_2t_2} \dot{X}_{t_2} E^{j_3t_3} \dot{X}_{t_3} E^{j_4t_4} \dot{X}_{t_4}\ddot{X}^{(k} \nn\\
& & \(h_1+h_4\) \ddot{X}_{j_2}[X^{i)},X_{j_3}][X_{j_4},X_j] \nn \\
& & +\frac{1}{12} E^{j_1t_1} \ddot{X}_{t_1} E^{j_2t_2} \ddot{X}_{t_2}\left(1+\dot{X}_u E^{uv}\dot{X}_v\right)^2 E^{j_3t_3} \dot{X}_{t_3} \ddot{X}^{(k} \nn\\
& & \Big( h_1 [X^{i)},\dot{X}_{j_1}][X_{j_3},X_{j_2}]
+h_2 [\dot{X}^{i)},X_{j_1}][X_{j_2},X_{j_3}]
+(h_3-h_4) [X^{i)},X_{j_1}][X_{j_2},\dot{X}_{j_3}] \nn \\
& & +(-h_3+h_5) [X^{i)},X_{j_1}][\dot{X}_{j_2},X_{j_3}]
+(h_4+h_5) [X^{i)},X_{j_3}][\dot{X}_{j_1},X_{j_2}] \Big) \nn \\
& & +\frac{1}{12} E^{j_1t_1} \ddot{X}_{t_1} E^{j_2t_2} \ddot{X}_{t_2}\left(1+\dot{X}_u E^{uv}\dot{X}_v\right)^2 E^{j_3t_3} \dot{X}_{t_4} E^{j_4t_4} \dot{X}_{t_4}\ddot{X}^{(k} \nn\\
& & \Big( (2h_1+h_3+h_4) \ddot{X}_{j_3} [X^{i)},X_{j_1}][X_{j_4},X_{j_2}]
+h_2 \ddot{X}^{i)} [X_{j_3},X_{j_1}][X_{j_4},X_{j_2}] \nn \\
& & +(-h_1-h_4+h_5) \ddot{X}_{j_1} [X_{j_3},X_{j_2}][X_{j_4},X^{i)}] \Big)\nn \\
\lefteqn{R_1^{(ik);i_1} =} \nn \\
& & \frac{1}{12} E^{jt} \dot{X}_t V^{ik;i_1}{}_j
+\frac{1}{12} E^{j_1t_1} \dot{X}_{t_1} E^{j_2t_2} \dot{X}_{t_2}\Big( \dddot{X}\!{}^{(i}[\dot{X}^{k)},X_{j_1}][{X}^{i_1},X_{j_2}] \nn \\
& & + \ddot{X}^{(i}[\dot{X}^{k)},\dot{X}_{j_1}][X^{i_1},X_{j_2}] + \ddot{X}^{(i}[\dot{X}^{k)},X_{j_1}][{X}^{i_1},\dot{X}_{j_2}]
+\ddot{X}_{j_1}[\dot{X}^{(i},X_{j_2}][X^{|i_1|},\dot{X}^{k)}] \nn \\
& & + \frac{1}{2} \ddot{X}^{i_1}[\dot{X}^{(i},X_{j_1}][\dot{X}^{k)},X_{j_2}] \Big) \nn \\
& & +\frac{1}{6} E^{j_1t_1} \dot{X}_{t_1} E^{j_2t_2} \dot{X}_{t_2} E^{j_3t_3} \dot{X}_{t_3} \ddot{X^{(i}} \ddot{X}_{j_1} [\dot{X}^{k)},X_{j_2}][X^{i_1},X_{j_3}] \nn \\
& & -\frac{1}{6} E^{jt} \ddot{X}_t \left(1+\dot{X}_u E^{uv}\dot{X}_v\right) E_{j_2t_2} \dot{X}^{t_2} \ddot{X}^{(k} \nn \\
& & \Big(h_1 [X^{i)},\dot{X}^{(j_2}][X^{i_1)},X_j]+h_2 [\dot{X}^{i)},X^{(j_2}][X_j,X^{i_1)}] \nn \\
& & +h_3 [X^{i)},X_j][X^{(j_2},\dot{X}^{i_1)}]
+h_4 [X^{i)},X^{(j_2}][\dot{X}^{i_1)},X_{j}] + h_5 [X^{i)},X^{(j_2}][\dot{X}_{j},X^{i_1)}]\Big)\nn \\
& & - \frac{1}{6} E^{jt} \ddot{X}_t \left(1+\dot{X}_u E^{uv}\dot{X}_v\right) E^{j_2t_2} \dot{X}_{t_2} E_{j_3t_3} \dot{X}^{t_3} \ddot{X}^{(k} \nn \\
& & \Big( h_1 [X^{i)},X_{j_2}][X^{(j_3},X_j]\ddot{X}^{i_1)}
+ \frac{1}{2} h_2 \ddot{X}^{i)}[X^{i_1},X_{j_2}][X^{j_3},X_j] \nn \\
& & \frac{1}{2} h_3 [X^{i)},X_j][X^{i_1},X_{j_2}]\ddot{X}^{j_3}
+h_4 [X^{i)},X^{(j_3}][X_{j_2},X_j]\ddot{X}^{i_1)} \nn \\
& & +\frac{1}{2} h_5 \ddot{X}_j[X^{i)},X^{j_3}][X_{j_2},X^{i_1}]\Big), \nn \\
\lefteqn{R_0^{(ikm)} = }\nn\\
& & \frac{1}{12} E^{jt} \dot{X}_t Q^{ikm}{}_j
+\frac{1}{24} E^{j_1t_1} \dot{X}_{t_1} E^{j_2t_2} \dot{X}_{t_2}\Big( \dddot{X}\!{}^{(i}[\dot{X}^{k},X_{j_1}][\dot{X}^{m)},X_{j_2}] \nn \\
& & +2\ddot{X}^{(i}[\dot{X}^{k},\dot{X}_{j_1}][\dot{X}^{m)},X_{j_2}]\Big)\nn \\
& & +\frac{1}{12} E^{j_1t_1} \dot{X}_{t_1} E^{j_2t_2} \dot{X}_{t_2} E^{j_3t_3} \dot{X}_{t_3} \ddot{X^{(i}} \ddot{X}_{j_1} [\dot{X}^{k},X_{j_2}][\dot{X}^{m)},X_{j_3}] \nn \\
& & +\frac{1}{12} E^{jt} \ddot{X}_t \left(1+\dot{X}_u E^{uv}\dot{X}_v\right) E^{j_2t_2} \dot{X}_{t_2} E^{j_3t_3} \dot{X}_{t_3} \ddot{X}^{(k} \ddot{X}^{m}\nn \\
& & \Big( h_1 [X^{i)},\dot{X}_{j_2}][X_{j_3},X_j]+h_2 [X_{j_2},\dot{X}^{i)}][X_{j_3},X_j] \nn \\
& & +h_3 [X_{j_2},\dot{X}_{j_3}][X^{i)},X_j]
 +h_4 [X_{j},\dot{X}_{j_2}][X_{j_3},X^{i)}] + h_5 [X_{j_2},\dot{X}_{j}][X_{j_3},X^{i)}]\Big)\nn \\
& & +\frac{1}{12} E^{jt} \ddot{X}_t \left(1+\dot{X}_u E^{uv}\dot{X}_v\right) E^{j_2t_2} \dot{X}_{t_2} E^{j_3t_3} \dot{X}_{t_3} E^{j_4t_4} \dot{X}_{t_4}\ddot{X}^{(k} \ddot{X}^{m}\nn\\
& & \(h_1+h_4\) \ddot{X}_{j_2}[X^{i)},X_{j_3}][X_{j_4},X_j].
\eea
where $f_1,f_2,f_3,f_4,f_5,h_1,h_2,h_3$ are the same coefficients as in (\ref{currentsol}).
Since there is now no need to satisfy current conservation, we can put them all to zero.

\end{document}